\documentclass[12pt]{article}
\usepackage[preprint]{aasfix}
\usepackage{chicago}
\usepackage{xspace}
\usepackage{macros}

\lefthead{Pinto \& Eastman}
\righthead{Supernova Ia Lightcurve Physics}

\begin{document}

\title{The Physics of Type Ia Supernova Lightcurves:\\
I. Opacity and Diffusion}

\author{Philip A. Pinto\altaffilmark{1}}
\affil{Steward Observatory\\
University of Arizona\\Tucson, AZ 85721 USA}
\and
\author{Ronald G. Eastman}
\affil{General Studies Group\\
Lawrence Livermore National Laboratory\\
Livermore, CA 94550 USA}
\altaffiltext{1}{General Studies Group, Lawrence Livermore National 
Laboratory, Livermore CA 94550 USA}

\slugcomment{Submitted to The Astrophysical Journal}

\setcounter{footnote}{0}

\begin{abstract}
We examine the basic physics of type Ia supernova (\sneia) light curves with
a view toward interpreting the relations between peak luminosity, peak
width, and late-time slope in terms of the properties of the underlying
explosion models. We develop an analytic solution of the radiation
transport problem in the dynamic diffusion regime and show that it
reproduces bolometric light curves produced by more detailed calculations
under the assumption of a constant extinction coefficient.  This model is
used to derive the thermal conditions in the interior of \sneia, to address
the issue of time dependence in the modeling of \sneia light curves and
spectra and to show that these are intrinsically time-dependent phenomena.
The analytic model is then used to study the sensitivity of light curves to
various properties of supernova explosions. We show that the dominant
opacity arises from line transitions and discuss the nature of line
opacities in expanding media and the appropriateness of various mean
opacities used in light curve calculations.
\end{abstract}


\newpage

\section{Introduction}

The bolometric light curve is the simplest and most direct manifestation of
type Ia supernov\ae\ (hereinafter \sneia). For many years it had been
assumed that all Type Ia supernov\ae\ (hereinafter \sneia) were identical
explosions, with identical light curve shapes and peak luminosities
(c.f. \citeNP{WoosleyW86}). While evidence for this uniformity in the data
was never terribly convincing, the use of \sneia as the primary ``standard
candles'' for cosmological distance measurement provided a powerful
incentive for assuming this homogeneity.  This in turn lead naturally to a
search for an explosion model which might produce identical displays from
the diversity of progenitors supplied by stellar evolution.

It became clear from the light curve's rapid evolution that a relatively low
mass object was involved, one with a short radiative diffusion time
\cite{Arnett82}. The result of this search is 
what one might call the present ``standard model'', the
thermonuclear incineration of a carbon-oxygen white dwarf at the
Chandrasekhar mass (see \citeNP{WoosleyW86},
\citeNP{Arnett96} for details of this search and a review of various
models). The Chandrasekhar mass provides a point of convergent evolution
for various progenitor systems, offering a natural explanation of the
assumed uniformity of display. The high densities attained at the centers
of these objects provides as well a mechanism (as ill-defined as it may be
at present) for their ignition.

With the coming of age of various supernova searches,
(\citeNP{Hamuyetal93}, \citeNP{Pollas94}, \citeNP{EvansVM89},
\citeNP{BarbonBCPT93}) there has recently been an explosion in the
availability of high-quality data. It is now generally recognized that SNe
Ia exhibit a variety of light curve shapes, peak luminosities, and
maximum-light spectra. Perhaps most significant has been the discovery of
various regularities in the light curve data, the most famous of which we
will call the ``Phillips Relation'' (hereinafter PR) \cite{Phillips93}; the
brightest supernov\ae\ have the broadest light curve peaks. There is also
significant evidence (\citeNP{VaccaL96a}) that the decline at late times
is more gradual in the brighter supernov\ae.

This additional information provides new clues to the nature of these
explosions. We develop here, from first principles, a theoretical framework
for examining the formation of SNa Ia light curves and extracting underlying
properties of the explosions. While there have been a number of studies of
the light curve problem in \sneia (\citeNP{HoflichKW95} and references
therein, \citeNP{WeaverAW80}), none has included a detailed examination of
the physics of the opacity or radiation transport in these explosions.

We divide the present investigations into four parts. In the first we
derive an analytic solution for the co-moving frame transfer equation
in homologous expansion. We use this solution to examine the relative
importance of various terms in the equations and the suitability of various
approximations which have appeared in the literature. The analytic solution
also provides an important check on the accuracy of our subsequent
numerical solutions.

In the second part we examine the nature of the opacity. \sneia differ
significantly from other astrophysical objects in their composition; they
are entirely constituted of heavy elements. This allows a number of
interesting effects, present at a low level in all objects, to take on a
dominant role. We find that the extreme complexity of the atomic physics
allows for a transport of energy downward in frequency which has a profound
effect upon the radiative diffusion time. This transport also allows for
a process, akin to thermalization but not mediated by collisions, which
can lead to a spectrum which appears thermal but which need have little to
do with the gas temperature anywhere in the object.

In the third part we present several light curves calculated by solving the
time-dependent, multi-frequency transport equations. We use these detailed
solutions to examine the behavior of the mean opacity, and find that the
most natural {\it a priori} choice, the Rosseland mean, significantly
underestimates the true flux mean by up to a factor of five. The flux mean
appears to be, at least in LTE, remarkably constant both in time and in
radius.

Finally, we exploit the constancy of the flux mean to examine the effect of
varying the bulk parameters of the explosion on the shape of the light curve.
For present purposes, \sneia can be characterized by specifying density and
composition as a function of ejecta velocity. We will ignore, at least to
begin with, how this structure was brought about. We will also ignore any
departures from spherical symmetry---as will be seen, the problem is vexing
enough in one spatial dimension. We note however that Wang et
al.~\citeyear{Wang96} found no evidence for polarization in their
study of three SNe~Ia, nor has there been any other evidence from
optical polarimetry for gross asphericity in SNe~Ia explosions.

The post-explosion dynamics, which determine the density and velocity
structure, are quite simple. As in all strong point explosions, the
expansion becomes homologous after a time which is short compared with
the bolometric rise-time, with a velocity gradient everywhere equal to
the reciprocal of the elapsed time. The density structure is quite
smooth, with a density profile not very different from $\rho\propto
\exp(-4 v/v_0)$ (with $v_0$ some typical velocity $\propto
(E/M)^{1/2}$). There is usually a small amount of high-velocity
material at the surface in which the density drops more rapidly; as
this material is largely transparent long before the supernova is
observed we may ignore this detail. To a surprising degree, all
explosion models to date have this relatively simple structure. Thus,
the dynamics can be specified by the explosion energy and the mass of
the ejecta (in most models the progenitor is completely disrupted).

The other defining attributes of the explosion are in the composition of
the ejecta. Since it is radioactive nickel which leads to any optical
display at all, the amount of \nifsx and the depth to which it is buried in
the ejecta will obviously affect the light curve. The composition
also affects the opacity, obviously a determining parameter in a problem
concerning the escape of radiation.

We will thus define a supernova by its total mass, explosion energy, \nifsx
mass and opacity, possibly allowing for variations in the spatial
distribution of the latter two (we will find that this is relatively
unimportant). With a simple means of producing a light curve from these four
parameters, one may turn the problem around and use the light curve model to
extract values for these parameters from observations of \sneia.

We find that, with the exception of the total mass, variations in any
of these basic parameters lead to a behavior of the light curve which
is in the {\em opposite sense} of the PR. Varying the total mass,
however, can lead to a sequence of light curves in which the PR
behavior is reproduced.  This is by no means a proof that the full
richness of \snia light curve behavior cannot be obtained from
Chandrasekhar mass explosions.  It does however suggest that the total
mass of the explosion may be a natural and simple explanation for
observations.

This is the first paper of a series. The next paper \cite{PintoE96b}
explores the lightcurve behavior of simple models for \sneia.  Subsequent
papers will systematically explore the light curve and spectrum properties
of specific models for Type Ia supernovae.

\section{A Schematic Type Ia Supernova}\label{AnLCMod}

In this section we develop a simple analytic model for the thermal
evolution and light curve of a \snia. This will prove useful for estimating
physical conditions in the ejecta at various times after explosion, and for
illustrating the effect which changes in opacity, mass, energy deposition,
and explosion energy have on the bolometric light curve.

The ejecta of \sneia form an opaque, expanding sphere into which energy is
deposited by radioactive decay at an exponentially-declining rate. Because
the sphere is initially so opaque, this energy is converted into kinetic
energy of expansion on a hydrodynamic timescale\footnote{For a point
explosion like a SN Ia, the hydrodynamic timescale is comparable to the
elapsed time.}. At the earliest stages the ejecta is so optically thick
that the time it takes radiation to diffuse out is much longer than the
elapsed time.  The luminosity is therefore initially quite small.  As time
passes, the ejecta become more dilute and the diffusion time drops below
the (ever-increasing) elapsed time. Since the rate of energy input declines
exponentially with time, there is a peak in the light curve as soon as the
injected energy has an appreciable chance to escape conversion to kinetic
energy---when the diffusion time becomes comparable to the elapsed
time. While the fraction of deposited energy which escapes conversion will
continue to increase, this is more than offset by the decreasing energy
deposition rate.

Shortly after this peak, there will be a considerable amount of radiation
still trapped and diffusing outward. Since the energy deposition rate is so
rapidly declining, the luminosity will, for a time, exceed the rate of
deposition until the supernova empties itself of this excess stored
energy. Finally, as the rate of energy deposition, now from cobalt decay,
declines more slowly and the diffusion time becomes small, the luminosity
becomes equal to the instantaneous deposition rate.  There are thus two
milestones in the light curve. The first occurs near peak when the
luminosity first rises above the rate of energy deposition. The second
occurs when the excess, stored energy is exhausted and the luminosity falls
to equal the instantaneous deposition.  The elapsed time and the rate of
deposition are easily determined. The first is obvious and the second comes
from the decay of \nifsx to \fefsx and the transport of $\gamma$-rays --
fairly simple physics. Determining the diffusion time is a far more complex
matter, and most of the difficulty in producing synthetic light curves and
spectra arises from correctly characterizing the transport and escape of
thermalized radiation.

Arnett showed in two elegant papers
(\citeyearNP{Arnett80}; \citeyearNP{Arnett82}) that the ideas expressed
above could be demonstrated by a simple analytic model which accounts for
the deposition and escape of radiation from the expanding ejecta.  This
model predicted a bolometric light curve which was generically in good
agreement with observed Sn~Ia behavior.  Starting from the thermodynamics
of the trapped radiation, he showed that the luminosity at peak was equal
to the instantaneous energy deposition rate under the assumption of
constant opacity, and thus the first milestone occurs near peak bolometric
luminosity. A number of assumptions were made which, for a first attempt,
were quite reasonable, but which rendered suspect the precise predictions
for any particular model explosion. These included a constant density
structure, a constant opacity in both space and time, and the requirement
that the radial distribution of the energy deposition was identical to that
of the thermal energy.  Thus, while he could vary the expansion velocity
and the total mass, the effects of varying the structure of the ejecta and
of a realistic energy deposition profile were beyond examination.  In
mathematical terms, Arnett's solution was an eigenfunction expansion from
which only the fundamental mode is retained. We shall have more to say on
this later.

We take a no-frills approach similar to Arnett's \citeyear{Arnett82}, while
relaxing some of his more limiting assumptions so as to be able to address
additional questions such as how the density structure and distribution of
radioactive isotopes are manifested in the bolometric light curve.

We start by writing down the first two frequency-integrated moments of
the time dependent, co-moving frame radiative transport equation in
spherical geometry. The first moment equation, for the radiation
energy density, can be written, correct to all terms O(v/c), as
(cf. \citeNP{MihalasM84})
\begin{equation}
{{DE}\over{Dt}} + {1\over{r^2}}{{\partial}\over{\partial r}}
\left(r^2 F\right) + {v\over r}\left( 3E-P \right)
+ {{\partial v}\over {\partial r}}\left(E+P\right)
= \int_0^\infty\left[4\pi \eta_\nu - c\chi_\nu E_\nu\right]d\nu.
\label{rade}
\end{equation}
The second frequency-integrated moment, for the radiation momentum, is
\begin{equation}
{1\over{c^2}}{{DF}\over{Dt}} + {{\partial P}\over{\partial r}}
+ {{3P-E}\over r} + {2\over{c^2}}\left({{\partial v}\over {\partial
r}}
+ {v \over r}\right) F = -{1 \over c} \int_0^\infty \chi_\nu F_\nu d\nu.
\label{radm}
\end{equation}
Here $E$, $F$, and $P$ are the zero, first, and second
frequency-integrated moments of the radiation field: the energy
density, the flux, and the (isotropic) radiation pressure. $\chi_\nu$
is the extinction coefficient, and $\eta_\nu$ is the volume
emissivity. These are formidable equations to solve directly
(cf. \citeNP{EastmanP93} among others) and our goal here is to obtain
a simple and approximate solution.

The first and most important approximation we will employ is that the
expansion is homologous. As already described, SNe Ia are strong point
explosions; homologous expansion will be an excellent approximation if
the energy released by \nifsx decay does not strongly affect the
dynamics of the expansion. The energy available from \nifsx decay is
$3\times10^{16}$\ergg. This corresponds to the kinetic energy of a
gram of material traveling at nearly 2500 \kms, or, equivalently, a
velocity increment of the same magnitude over the velocity initially
imparted by the explosion. The significantly greater decay energy
available from decay all the way to \fefsx is less relevant as most of
the \cofsx decay energy is emitted at times when the supernova is
becoming optically thin. Since the observed expansion velocity of SNe
Ia is in excess of $10^4$ \kms, we expect that this additional source
of energy will have a modest, but perhaps not completely negligible,
effect upon the velocity structure. Furthermore, as the time to
maximum light, $t_{max}$, is observed to be at least twice as large
the \nifsx\ decay time, most of the hydrodynamic effect of \nifsx\
decay will have occured prior to a supernova's discovery. If we take
the ejecta's density structure from an explosion calculation which has
allowed this additional energy to accelerate the ejecta for the first
few days we will have taken this effect sufficiently into account.  We
will therefore take the outer edge of the supernova, or at least of a
fiducial mass shell which contains virtually all of the mass, to be at
a velocity $v_{max}$ and a radius
\begin{equation}
R(t) = R_0 + v_{max} t
\label{Rt}
\end{equation}
where $R_0$ is the initial radius of the progenitor.  For this type of
expansion law there is an associated time scale
\begin{equation}
t_{sc} = R_0/v_{max}.
\label{tsc}
\end{equation}
which will be one of the parameters of the solution.

A major simplification is the so-called {\it Eddington Approximation},
wherein the radiation field is everywhere isotropic: $E =
3P$. This is certainly valid during the early, optically thick stages
of evolution, but breaks down when the ejecta become
transparent. The error that this assumption introduces at late times lies
in the energy distribution, and has much less effect on the
bolometric luminosity which is the only link this simple analytic
model has to SN~Ia observations. We note in passing that this assumption
has dire consequences for the calculation of the energy deposition,
where the deposition rate is proportional to the gamma-ray energy density.

At the temperatures and densities of maximum light SNe~Ia, the gas
energy density is less than the radiation field energy density by
a large factor.  The radiation field energy density is
\begin{equation}
\rho e_{rad}\equiv aT^4\sim 1210 \left({T\over2\times10^4\ \K}\right)^4
\end{equation}
which greatly exceeds both the thermal kinetic energy density
\begin{equation}
\rho e_{kin} \equiv {3\rho N_A \over 2 A}(i+1) k T \sim
0.4\left({\rho\over 10^{-12}{\rm }g~{\rm cm}^{-3}}\right)
\left({T\over2\times10^4~K}\right)
\left({A\over56}\right)
\end{equation}
and the ionization energy density
\begin{equation}
\rho e_{ion}\equiv {\rho N_A\over A}\langle E\rangle
\sim 
0.5\left({\rho\over 10^{-12}}{\rm }g~{\rm cm}^{-3}\right)
\left({\langle E\rangle\over 30\ \hbox{eV}}\right)
\left({A\over56}\right)
\end{equation}
where $N_A$ is Avogadro's number, $A$ is the mean mass per nucleon,
$i$ is the average ionization,
and $\langle E\rangle$ is the mean ionization energy.

The dominance of radiation over internal energy 
permits us to ignore the gas internal energy and set
\begin{equation}
\int_0^\infty\left[4\pi \eta_\nu - c\chi_\nu E_\nu\right]d\nu = \epsilon
\label{ThermBal}
\end{equation}
where $\epsilon$ is the volume rate of $\gamma$-ray deposition.
Equation~\ref{ThermBal} is equivalent to saying that as soon as high-energy
radiation from decay is absorbed, it is immediately re-radiated as thermal
emission. The mechanism by which this happens is collisional; $\gamma$-rays
Compton scatter, producing high-energy electrons which then rapidly
transfer their energy to the plasma. This occurs on a timescale which is
short compared with any other timescale important to the problem of energy
transport.

Much of the radiation transport in the peak-light phase of the light curve
occurs as diffusion. As discussed earlier, the diffusion time for radiation
to escape the supernova is, at first, long compared with the hydrodynamical
(elapsed) time. This puts us in the {\em dynamic diffusion} regime of
\cite{MihalasM84}. Following the discussions referenced therein, it is
important when in this regime to retain {\em all} of the terms in the
radiation energy equation to O(v/c) . However, in the frequency integrated
radiation momentum equation, equation~\ref{radm}, it is appropriate to discard all
time- and velocity-dependent terms. This difference in treatment is
intuitively evident when one realizes that we are vitally interested in
determining the energy density and its flow within the supernova, yet the
radiation momentum has little effect upon the supernova's dynamics after
the first few days. The radiation momentum equation is thus reduced to the
familiar diffusion form
\begin{equation}
F = - {c \over{3 \chi}} {{\partial E}\over{\partial r}}.
\label{fluxdef}
\end{equation}
where $\chi$ is an appropriately defined frequency-averaged mean opacity
(see section~\ref{MeanOpac} below).  Using this and the previous
approximations in the radiation energy equation, the transport equation
becomes
\begin{equation}
{{D E}\over{D t}} -
{c\over{3r^2}}{{\partial}\over{\partial r}}\left({{r^2}\over{\chi}}
{{\partial E}\over{\partial r}}\right)
+ {{4\dot{R}}\over{R}}E = \epsilon.
\label{primaryI}
\end{equation}
In general, equation \ref{primaryI} is too complicated to solve
analytically. However, for certain conditions specified below the spatial
and temporal parts of the solution may be separated, and equation
equation~\ref{primaryI} reduced to two ordinary differential equations. 

It is convenient to homologously scale all of the remaining quantities
in terms of the time $t$ and $x = r/R$, the (dimensionless) fractional
radius. Since the gas is radiation dominated, $E\propto R^{-4}\propto
(1+t/t_{sc})^{-4}$, we write
\begin{eqnarray}
E(x,t) & = & {\cal E}(x,t) E_0 \left({{R_0}\over{R(t)}}\right)^4 \nonumber\\
       & = & E_0 \psi(x) \phi(t) \left({{R_0}\over{R(t)}}\right)^4,
\label{Ehom}
\end{eqnarray}
where $\psi(x)$ describes the radial variation, $\phi(t)$ the temporal
variation, and $E_0$ is the initial energy density.

The density can be written as
\begin{equation}
\rho(r,t) = \rho_0 \tilde{\rho}(x) \left( {{R_0}\over{R(t)}}\right)^3,
\label{Rhohom}
\end{equation}
where $\tilde\rho(x)$ is the radial profile of the density, normalized so
that $\tilde\rho(0)=1$.

The extinction coefficient $\chi$ is the mass opacity coefficient
$\kappa$ times the density. We allow $\kappa$ to have an intrinsic 
radial dependence, described by $\tilde{\kappa}(x)$, as well as a time
dependence, $\zeta(t)$:
\begin{equation}
\chi(r,t) = \kappa_0 \tilde{\kappa}(x) \zeta(t) \rho_0 \tilde{\rho}(x)
\left({{R_0}\over{R(t)}}\right)^3.
\label{Chihom}
\end{equation}
Separation of variables is possible only if the opacity does not
depend upon the energy density (i.e. the temperature).  Remarkably,
conditions may actually conspire to produce a mean opacity which is
roughly constant with both time and depth through the ejecta
(section~\ref{MeanOpac}).

The volume energy deposition rate $\epsilon$ will scale as
\begin{equation}
\epsilon(r,t) = {{3 M_{ni} \epsilon_0}\over{4\pi R_0^3 }}
\theta(t) \Lambda(x,t) \left({{R_0}\over{R(t)}}\right)^3
\label{eps}
\end{equation}
where $\epsilon_0=E_{Ni}+E_{Co}=1.73\ \,\MeV + 3.69\, \MeV=5.42\, \MeV$
is the total energy available from decay, per atom of \nifsx, and
$\Lambda(x,t)$ is the dimensionless energy deposition function which
results from $\gamma$-ray transport and escape.
The total production rate of decay energy as a function of $t$ is
described by $\theta(t)$, given as
\begin{equation}
\theta(t) = \epsilon_0^{-1}
 \left\{
E_{Ni} e^{-t/\tau_{Ni}} + 
{{E_{Co}\tau_{Co}}\over{\tau_{Ni}+\tau_{Co}}}
\left(e^{-t/\tau_{Co}} - e^{-t/\tau_{Ni}}\right)\right\}.
\label{Thetadef}
\end{equation}
It is convenient to define the total energy available from the
\nifsx$\rightarrow$\fefsx decay chain (excluding neutrinos)
in terms of the total kinetic energy of the gas,
\begin{equation}
\tilde{\epsilon} =  {{3 M_{ni} \epsilon_0}\over{4\pi R_0^3 E_0 }},
\end{equation}
and the initial diffusion time from the center as
\begin{equation}
\tau_d = {{3 \chi_0 R_0^2}\over c},
\end{equation}
where $\chi_0 = \rho_0\kappa_0$.

Substituting equations~\ref{Ehom}, \ref{Rhohom}, \ref{Chihom}, \ref{eps}
and \ref{Thetadef} into equation~\ref{primaryI} gives
\begin{equation}
\tau_d\zeta(t) {{R_0}\over{R(t)}} \dot{\phi} \psi
- \phi {1\over{x^2}} {{\partial}\over{\partial x}}
\left( {{x^2}\over{\tilde{\chi}}}
       {{\partial \psi}\over{\partial x}}\right)
= \tau_d\tilde{\epsilon} \zeta(t)\theta(t)\Lambda(x,t)
\label{primary2}
\end{equation}
where we have combined the spatial shape of the density and opacities into
the function $\tilde{\chi}(x)=\tilde{\rho}\tilde{\kappa}$.
Equation~\ref{primary2} is the principle equation which describes the
evolution of the radiation field energy density. 

Next we must specify the boundary conditions.  At the center there is a
reflection boundary condition where the flux vanishes, or equivalently,
$\psi^\prime(0)=0$. At the surface we use a solution to the plane-parallel
grey atmosphere problem, assuming that the thickness of the surface layers
is small compared with the radius. This can be expressed as
\begin{equation}
\psi(x) = {3\over 4}\psi_e\left(\tau + {2\over3}\right)
\end{equation}
At the outer edge, $\tau=0$, and we have
\begin{equation}
\psi(1) = {1\over 2}\psi_e
\end{equation}
and thus
\begin{equation}
\psi(x) = {3\over 2}\psi(1)\left(\tau + {2\over3}\right)
\end{equation}
The boundary condition is then
\begin{equation}
\psi(1) = {2\over 3}\left.{{d\psi}\over{d x}}\right|_{x=1}
\left(\left.{{d \tau}\over{d x}}\right|_{x=1}\right)^{-1}.
\end{equation}
Since the optical depth to the surface from radius r is
\begin{eqnarray}
\tau &  = & - \int_r^{R(t)} \rho\kappa dr \nonumber \\
     &  = & -\rho_0 \kappa_0 R(t) \int_x^1
\tilde{\rho}(x^\prime)\tilde{\kappa}(x^\prime) dx^\prime
\end{eqnarray}
we have 
\begin{equation}
\left.{{d\tau}\over{d x}}\right|_{x=1} = - \rho_0 \kappa_0
R(t)\tilde{\rho}(1)\tilde{\kappa}(1).
\label{bc2}
\end{equation}
It is more convenient as a boundary condition to require the solution
to go to zero at some radius beyond $x=1$. Extrapolating
equation~\ref{bc2} linearly, we find that this is equivalent to
requiring $\psi(x_0) = 0$ at
\begin{equation}
x_0 = 1 - 2/3 \left.{{d x}\over{d\tau}}\right|_{x=1}.
\label{bcon}
\end{equation}
The value of $x_0$ increases as density declines and strictly speaking,
this will introduce a time dependence into the spatial solution which
violates the conditions making equation~\ref{primary2} separable.  However if we
consider this to be a slow, quasi-static change, the solution obtained by
separation of variables is not too inaccurate, and will be adequate for our
needs. We will touch on this again.

To solve equation~\ref{primary2} we follow the usual procedure for
separation of variables and first find a solution to the homogenous
equation, where energy deposition from decay is set to zero. In the
absence of any sources, equation~\ref{primary2} can be written as
\begin{equation}
 {{R_0}\over{R(t)}}\tau_d\zeta(t){\dot{\phi}\over{\phi}} 
=  {1\over{x^2\psi}}{{\partial}\over{\partial x}}
\left({{x^2}\over{\tilde{\chi}}}
{{\partial \psi}\over{\partial x}}\right).
\end{equation}
Since the terms on the left hand side of this equation depend only on $t$,
while those on the right hand side depend only on $x$, each must be equal
to a constant independent of either $x$ or $t$. Let this constant be
$\alpha$.  We can then write, for the spatial part,
\begin{equation}
{1\over{x^2\psi(x)}}{{\partial}\over{\partial x}}
\left({{x^2}\over{\tilde{\chi}}}
{{\partial \psi}\over{\partial x}}\right)=-\alpha.
\label{spatial}
\end{equation}
For $\tilde{\chi} = 1$, the solution can be written
\begin{equation}
\psi(x) = {{\sin(\alpha^{1/2} x/x_0)}\over{\alpha^{1/2} x/x_0}}
\end{equation}
where the eigenvalue $\alpha$ depends upon the total optical
depth. For large opacity, $d\tau/dx$ is large, and the boundary
condition equation~\ref{bc2} approaches the radiative-zero condition,
$\psi(1) = 0$.  For constant $\chi$, the radiative-zero eigenvalues
are $\alpha_n = n^2\pi^2$ and the eigenfunctions are
\begin{equation}
\psi_n(x) = \sqrt{2}{{\sin(n\pi x)}\over{x}}.
\end{equation}

Given $\alpha_n$, the temporal part of the solution, $\phi_n(t)$ is
determined by the homogenous equation
\begin{equation}
{{R_0}\over{R(t)}}\tau_d\zeta(t){\dot{\phi_n}\over{\phi_n}} 
=-\alpha_n.
\end{equation}
For $\zeta(t)=1$ (opacity does not change with time), the solution can be
written as
\begin{equation}
\phi_n(t)=\exp\left(-{\alpha_n t\over \tau_d}
\left(1+{t \over 2 t_{sc}}\right)\right).
\label{peakshape}
\end{equation}

When $\tilde{\chi}$ is an arbitrary function, there are no analytic
solutions to equation~\ref{spatial}, and we must determine the eigenfunctions
numerically. Eigenvalues are determined by a Rayleigh-Ritz procedure, and
a discrete representation of their corresponding eigenfunctions is obtained
by relaxation. The basin of convergence to a desired eigenfunction for this
process is surprisingly small; for most interesting $\tilde{\rho}(x)$,
eigenvalues must be determined to better than a percent for the resulting
relaxation to converge to the desired solution. We prefer to normalize the
solutions such that the functions $\psi_n$ are orthonormal with respect to
the inner product
\begin{equation}
<f|g>\equiv\int_0^1 f(x)g(x)x^2 dx.
\end{equation}

As the solution progresses in time, the spatial solution changes because
the value of $x_0$ increases with decreasing optical depth.  To avoid the
need of a new solution of equation~\ref{spatial} at each time, in much of the
discussion which follows, we will take the radiative zero solution.  This
allows a single set of eigenvalues to be used at all times. For more
realistic calculations, we note that the change in eigenvalues due to
changes in $\tilde{\chi}$ over time is slow and continuous. We may thus
continuously re-solve the eigenvalue problem as we evolve the solution in
time.

\begin{figure}[tp]
\postfig{l}{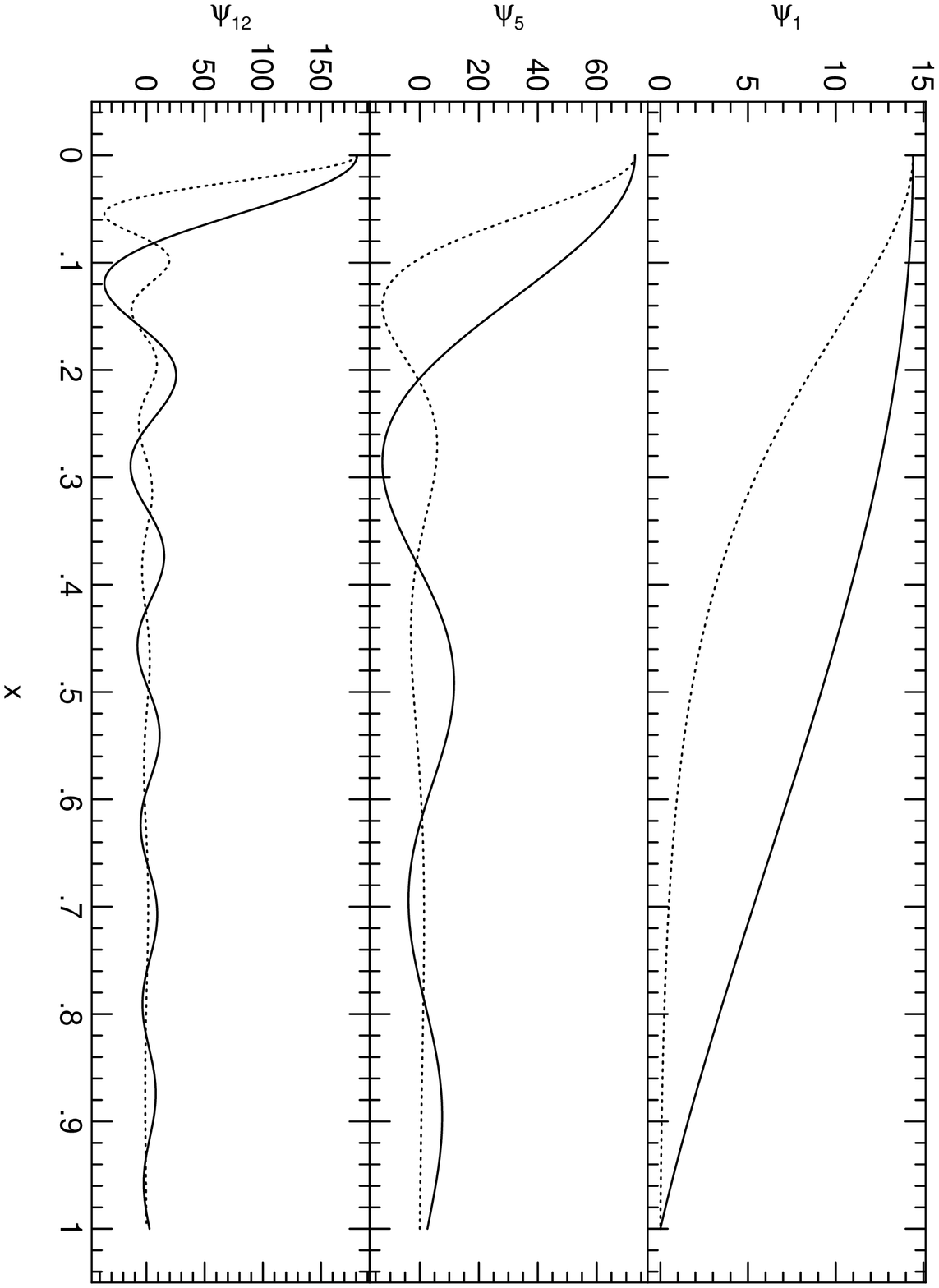}
\caption{Eigenfunctions for mode numbers 1, 2, 5, and 10 for exponential
density laws with constant $K=0,4$}
\label{eigenplot}
\end{figure}

As an example, Table \ref{eigenvalues} lists the first 25 eigenvalues for
the radiative zero solution with $\tilde{\chi}=e^{-kx}$, with $k=1$ to 4. A
selection of these functions is shown in Figure~\ref{eigenplot}.

\begin{table}
\centerline{
\begin{tabular}{rcccccc}
\multicolumn{7}{c}{\sc Table 1}      \\
\multicolumn{7}{c}{Eigenvalues for \protect$\rho(x) = e^{-kx}$}      \\
\hline
\hline
mode & k=0      &   k=0.25   & k=0.5      & k=1        & k=2        & k=4        \\
\hline
 1 & 9.8696$_0$ & 1.1878$_1$ & 1.4250$_1$ & 2.0298$_1$ & 3.9256$_1$ & 1.1637$_2$ \\
 2 & 3.9478$_1$ & 4.5567$_1$ & 5.2460$_1$ & 6.9007$_1$ & 1.1596$_2$ & 2.9175$_2$ \\
 3 & 8.8827$_1$ & 1.0152$_2$ & 1.1573$_2$ & 1.4922$_2$ & 2.4089$_2$ & 5.6453$_2$ \\
 4 & 1.5791$_2$ & 1.7978$_2$ & 2.0414$_2$ & 2.6115$_2$ & 4.1477$_2$ & 9.3937$_2$ \\
 5 & 2.4674$_2$ & 2.8037$_2$ & 3.1773$_2$ & 4.0488$_2$ & 6.3781$_2$ & 1.4181$_3$ \\
 6 & 3.5531$_2$ & 4.0328$_2$ & 4.5651$_2$ & 5.8043$_2$ & 9.1012$_2$ & 2.0015$_3$ \\
 7 & 4.8362$_2$ & 5.4852$_2$ & 6.2050$_2$ & 7.8782$_2$ & 1.2317$_3$ & 2.6900$_3$ \\
 8 & 6.3167$_2$ & 7.1610$_2$ & 8.0969$_2$ & 1.0271$_3$ & 1.6027$_3$ & 3.4837$_3$ \\
 9 & 7.9946$_2$ & 9.0602$_2$ & 1.0241$_3$ & 1.2982$_3$ & 2.0230$_3$ & 4.3828$_3$ \\
10 & 9.8700$_2$ & 1.1183$_3$ & 1.2637$_3$ & 1.6011$_3$ & 2.4927$_3$ & 5.3874$_3$ \\
11 & 1.1943$_3$ & 1.3529$_3$ & 1.5285$_3$ & 1.9360$_3$ & 3.0117$_3$ & 6.4974$_3$ \\
12 & 1.4213$_3$ & 1.6098$_3$ & 1.8186$_3$ & 2.3027$_3$ & 3.5802$_3$ & 7.7130$_3$ \\
13 & 1.6681$_3$ & 1.8891$_3$ & 2.1338$_3$ & 2.7013$_3$ & 4.1980$_3$ & 9.0342$_3$ \\
14 & 1.9346$_3$ & 2.1907$_3$ & 2.4743$_3$ & 3.1317$_3$ & 4.8653$_3$ & 1.0461$_4$ \\
15 & 2.2208$_3$ & 2.5147$_3$ & 2.8400$_3$ & 3.5941$_3$ & 5.5819$_3$ & 1.1993$_4$ \\
16 & 2.5269$_3$ & 2.8610$_3$ & 3.2309$_3$ & 4.0883$_3$ & 6.3480$_3$ & 1.3632$_4$ \\
17 & 2.8526$_3$ & 3.2297$_3$ & 3.6471$_3$ & 4.6144$_3$ & 7.1635$_3$ & 1.5375$_4$ \\
18 & 3.1981$_3$ & 3.6207$_3$ & 4.0885$_3$ & 5.1724$_3$ & 8.0284$_3$ & 1.7225$_4$ \\
19 & 3.5634$_3$ & 4.0341$_3$ & 4.5551$_3$ & 5.7624$_3$ & 8.9428$_3$ & 1.9180$_4$ \\
20 & 3.9484$_3$ & 4.4699$_3$ & 5.0469$_3$ & 6.3842$_3$ & 9.9066$_3$ & 2.1241$_4$ \\
21 & 4.3532$_3$ & 4.9279$_3$ & 5.5640$_3$ & 7.0379$_3$ & 1.0920$_4$ & 2.3408$_4$ \\
22 & 4.7778$_3$ & 5.4084$_3$ & 6.1063$_3$ & 7.7235$_3$ & 1.1983$_4$ & 2.5680$_4$ \\
23 & 5.2221$_3$ & 5.9112$_3$ & 6.6739$_3$ & 8.4410$_3$ & 1.3095$_4$ & 2.8059$_4$ \\
24 & 5.6862$_3$ & 6.4364$_3$ & 7.2667$_3$ & 9.1905$_3$ & 1.4256$_4$ & 3.0543$_4$ \\
25 & 6.1700$_3$ & 6.9840$_3$ & 7.8848$_3$ & 9.9719$_3$ & 1.5467$_4$ & 3.3133$_4$ \\
\\
ratio &  1.0    &   1.1322   &  1.2782    &   1.6163   &   2.5063   &    5.3643  \\
\hline
\end{tabular}  
}
\caption{ The first 25 eigenvalues for the spatial equation
\protect\ref{spatial} The final row is the ratio of subsequent eigenvalues
to $n^2\pi^2$, the asymptotic limit.
}
\label{eigenvalues}
\end{table}

Turning now to our original, inhomogeneous transport equation, equation
\ref{primary2}, the general solution for $E(x,t)$ is an expansion in terms
of the eigenfunctions $\psi_n$:
\begin{equation}
E(x,t) = \sum_{m=1}^\infty E_m \psi_m(x) \phi_m(t).
\label{E_expansion}
\end{equation}
If we substitute this expansion into the inhomogeneous equation
\ref{primary2}, multiply by $x^2 \psi_n(x)$ and integrate from $x=0$
to 1, we get
\begin{equation}
\dot{\phi}_n + {{\alpha_n}\over{\tau_d}} \left(1+t/t_{sc}\right) \zeta(t)^{-1} \phi_n
= \tilde{\epsilon}\theta(t)\left(1+t/t_{sc}\right)<\Lambda|\psi_n>.
\label{inhomog}
\end{equation}
where $<\Lambda|\psi_n>$ is the overlap integral of $\Lambda(x,t)$ with
eigenfunction $\psi_n(x)$.

For simple forms of $\zeta(t)$ and $<\Lambda|\psi_n>$, equation~\ref{inhomog}
can be integrated analytically. It is, however, a straightforward matter to
integrate this equation numerically, and one need not be limited by the
analytic integrability of these two functions.

We must now put back the dimensional constants to be able to calculate a
light curve. Starting with the definition of the flux equation~\ref{fluxdef}, we have
for the luminosity
\begin{equation}
L(t) = - {{4\pi c R_0 E_0}\over{3\chi_0}} {1\over{\zeta(t)}}
\sum_{n=0}^\infty \phi_n(t) \left( {{x^2}\over{\tilde{\chi}(x)}}
{{d\psi}\over{dx}}\right)_{x=1}.
\label{difflum0}
\end{equation}

Because the boundary conditions are $\psi(0) = 0$ and
$\psi^{\scriptstyle\prime}(x_0)=0$, there is no scale to the problem, and
we are free to impose a third condition on the overall normalization of the
solution.  We can compare the luminosity with the energy deposited at late
times by noting that when the timescale over which the energy deposition
$\theta(t)\Lambda(t,x)$ changes becomes long compared with the diffusion
time, the $\phi_n$ go asymptotically to
\begin{equation}
\phi_n = {{\tau_d}\over{\alpha_n}} \tilde{\epsilon}
\theta(t)\zeta(t)\langle\Lambda|\psi_n\rangle.
\label{asymphi}
\end{equation}
If we integrate equation~\ref{spatial} over volume, we find that
\begin{equation}
\left( {{x^2}\over{\tilde{\chi}(x)}}
{{d\psi}\over{dx}}\right)_{x=1} = -\alpha_n I_n(1),
\end{equation}
where the total radiation energy interior to $x$ is
\begin{equation}
I_n(x) = \int_0^{x} x^2 \psi_n dx.
\end{equation}
Putting these two results into the expression for the luminosity and using
the definitions of $\tau_d$ and $\tilde{\epsilon}$ gives
\begin{equation}
L(t) = 3 M_{Ni} \epsilon_0 \theta(t) \sum_{n=0}^\infty \phi_n(t)
\langle\Lambda|\psi_n\rangle I_n(1)
\end{equation}
If $\Lambda(x,t)$ is constant in $x$, we can let $\Lambda(x,t) \sim
\Lambda(t)$, and the sum becomes
\begin{equation}
\Lambda(t) \sum_{n=0}^\infty \left[\int_0^{x_s} x^2 \psi_n dx \right]^2.
\end{equation}

Requiring, then, that the solutions $\psi_n$ to be normalized such that
this is $\Lambda(t)/3$ leads finally to
\begin{equation}
L(t) = M_{Ni} \epsilon_0 \theta(t) \Lambda(t),
\end{equation}
the instantaneous energy deposition.

In order to examine the effect of including an increasing number of modes
on the {\em shape} of the light curve it is convenient to renormalize the
energy deposition such that the correct total amount of energy is deposited
per unit time into whatever modes are included in the calculation. We
therefore divide the energy deposition factor
$\langle\Lambda|\psi_n\rangle$ by the quantity
\begin{equation}
f = {{\sum_n \langle\Lambda|\psi_n\rangle}\over{\int_0^1 x^2 \Lambda dx}}.
\end{equation}
This has an aliasing effect of overestimating the power in the included
modes just enough to bring the deposited power to the correct value.

For the $\gamma$-ray deposition function, $\Lambda(x,t)$, we compute a
solution to the time independent $\gamma$-ray line transport problem at each
time $t$. It is not necessary to solve the fully time dependent transport
problem because the flight time for $\gamma$-rays before absorption or
escape is much smaller than any other time scale of interest. We have
performed this calculation two ways: in one case, each of the most
important lines in the \nifsx\ and \cofsx decay spectra are separately
transported, as described by \citeN{WoosleyEWP94}.  Alternately, we perform
the calculation for just two lines, one for \nifsx\ at the emission
weighted mean energy of 479~keV, and the other for \cofsx, at the emission
weighted mean energy of 1.13~Mev. The two methods give results which agree
with each other and to exact Monte Carlo results to better than a percent
over the first 30-40 days of the light curve.

\subsection{Comparison with a Multi-Group Calculation}

In order to assess the accuracy or realism of the analytic model it is
instructive to compare its predicted bolometric light curve with one
produced by a more detailed (and expensive) calculation.  We have therefore
used the procedures outlined in the last section to compute the light curve
of a model which approximates the main properties of the well studied \mch
deflagration Model W7 of \citeN{NomotoTY84}. For the analytic model we take
$M_{tot}=1.386$~\Msun, $M_{56}=0.625$~\Msun, $R_0=1.4\times10^8$~cm and
$v_{max}= 10^9$\cms. The peak light curve is insensitive to the choice of
initial temperature, and the value $10^{10}$~K was used. The density is
constant with radius, and a radiative zero boundary condition is
assumed. The mean opacity was taken to be the constant value
$\kappa_0=0.13$~cm$^2$~g$^{-1}$.  The abundance of \nifsx\ is unity out to
a radius given by $r_{56}=(M_{56}/M_{tot})^{1/3}(R_0+v_{max}t)$ and zero
beyond, and the $\gamma$-ray deposition was computed as described above.

\begin{figure}[tp]
\postfig{l}{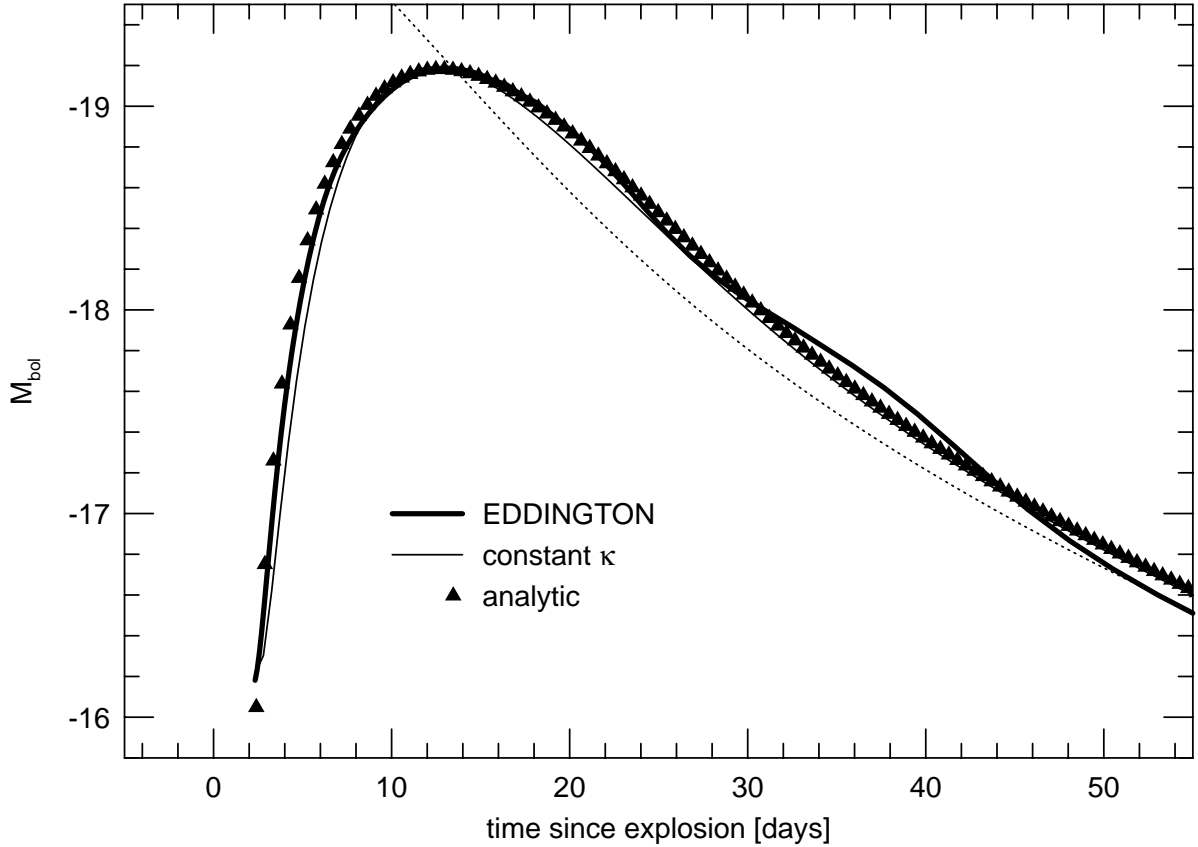}
\caption{A comparison of the bolometric light curve (solid line) of Model
W7 of \protect\citeN{NomotoTY84} as determined by a multi-group radiation
transport calculation performed with EDDINGTON, a numerical solution of the
grey transport equation for the same model employing a constant opacity,
and the analytic solution described in the text for a constant density
explosion of the same total mass, \nifsx mass, kinetic energy, and
opacity. The constant opacity calculations agree well with the
multi-group calculation with the exception of the secondary ``bump'' which
is produced by a decrease in the mean opacity, thus allowing the release of
stored energy on a shorter timescale than in the constant opacity
cases.}
\label{annedd}
\end{figure}

Figure (\ref{annedd}) compares the analytic version of ``W7'' with the
bolometric light curve obtained by a multi-group (3000 frequency points)
LTE transport calculation with EDDINGTON \cite{EastmanP93}.  The
EDDINGTON calculation used the actual structure and composition of Model
W7, and predicted a flux mean opacity (see section~\ref{MeanOpac}) close to
the value adopted for the grey calculation ($\kappa=0.13$~\cmsqrgm).  For
the first 30 days the agreement between the two calculations is excellent.
We note that the good agreement {\sl is} somewhat deceptive, because the
constancy of $\kappa$ with time arises by assumption in the grey model,
while no such assumption was made in the multi-group calculation. We will
show below, however, that there is some reason to expect that the opacity
will in fact be roughly constant.

The ``bump'' in the light curve of the EDDINGTON calculation which appears
at between 30 and 44~days reproduces similar features seen in the
observational data \cite{Suntzeff95}. It is caused by a decrease in the
mean opacity which allows stored energy to be released more quickly than in
the constant-opacity models. The constant opacity calculation lacks the
second bump, and falls onto the radioactive tail more slowly as a
result. The unrealistically short bolometric risetimes of these lightcurves
is due in part to the population III abundances employed in the explosion
model. We shall discuss this further below.

\subsection{Thermal Conditions In a Maximum Light SNe~Ia and Parameter
Sensitivity}

\begin{figure}[tp]
\postfig{p}{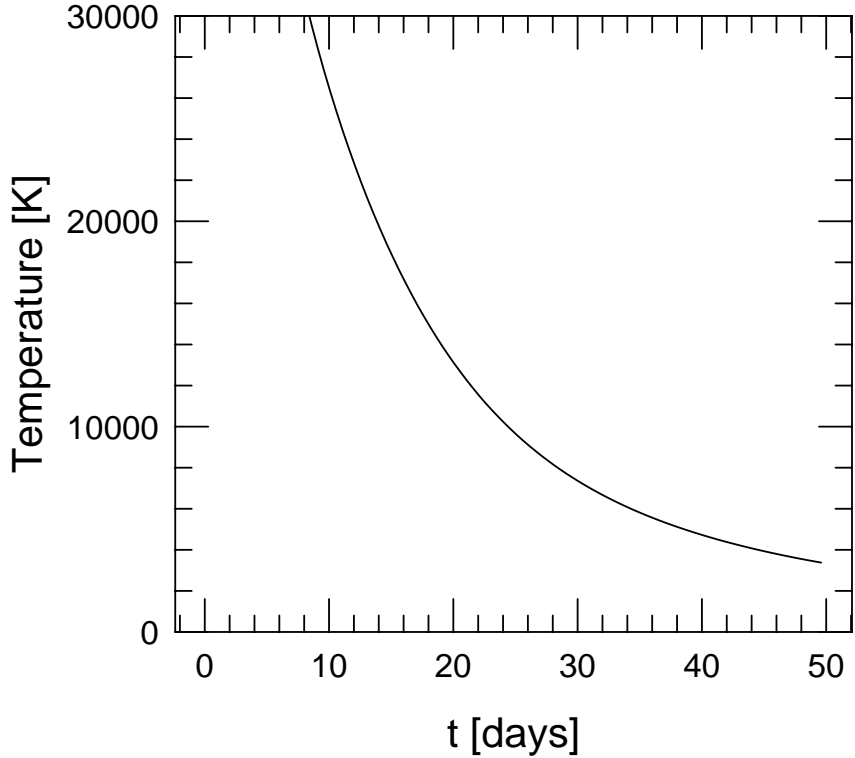}
\caption{Analytic model temperature solution at $x=0$ for a constant
density model having the same mass, kinetic energy and \nifsx\
mass as Model~W7.}
\label{analW7Temp}
\end{figure}

One application of the analytic model is to estimate temperatures in
SNe~Ia.  Figure~\ref{analW7Temp} shows central temperature versus time
for the ``W7''-like analytic model previously shown in
Figure~\ref{annedd}. At times $t<20$~days the central
temperature is $T(x=0)>13,000$~K, which puts the peak of the blackbody
spectrum at $\lambda_{Wein}\ltsim 2200$\AA, a wavelength at
which the optical depth due to lines is very large (section~\ref{opacity}).

\begin{figure}[tp]
\postfig{l}{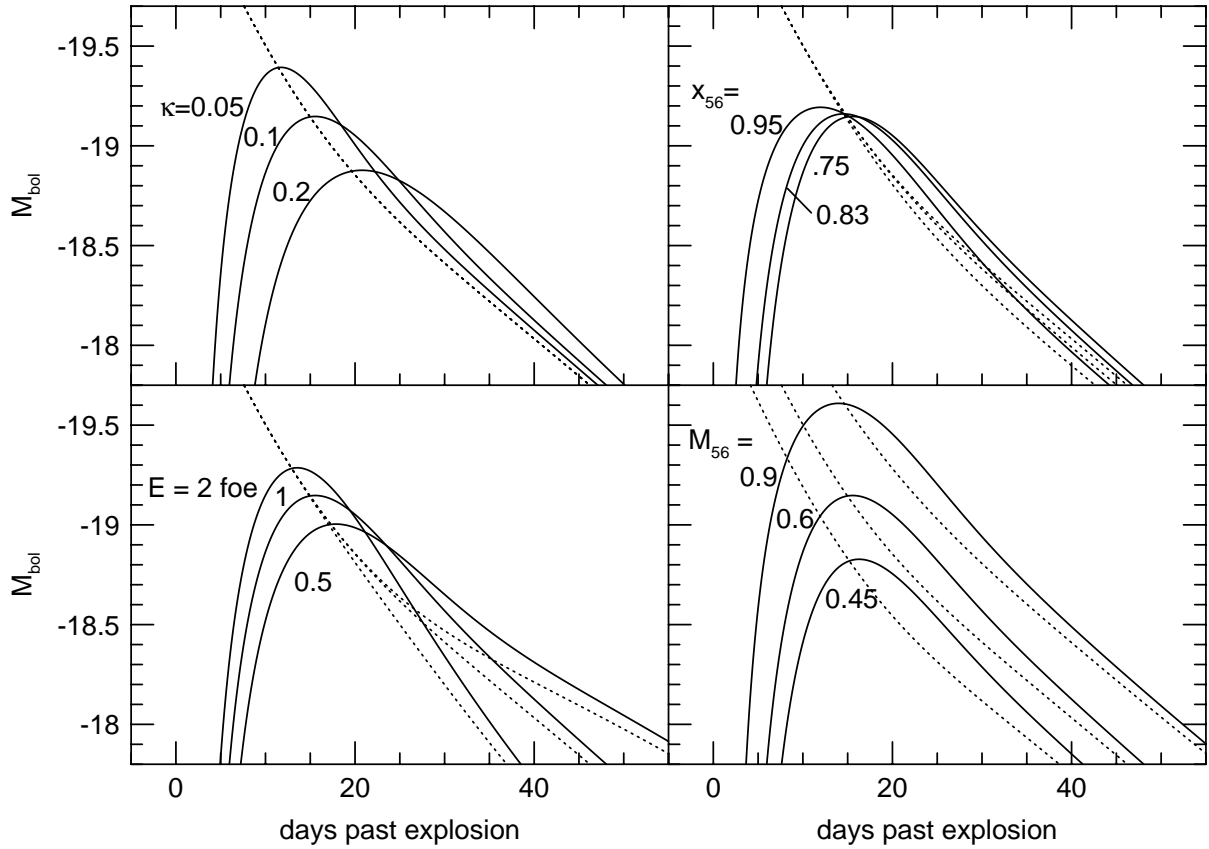}
\caption{The effect of varying the opacity, extent of deposition, and
explosion energy on the standard explosion (the center curve in each plot).
The instantaneous energy deposition rate is shown as a dotted line.}
\label{fourparamplot}
\end{figure}

Figure \ref{fourparamplot} demonstrates the dependence of the analytic
model's light curve solution upon changes in opacity, the distribution
of \nifsx, the mass of \nifsx, and the explosion energy, all for
Chandrasekhar-mass explosions. In all these calculations the fiducial
model is the same---the ``W7''-like model discussed in the previous
section.

In the first panel of Figure~\ref{fourparamplot}, the opacity is
varied by a factor of two above and below our fiducial model. The
effect is just what one would expect. A lower opacity decreases the
diffusion time, allowing radiation to escape earlier.  Spending less
time in the expanding, optically thick enclosure, the radiation
suffers a smaller loss to expansion. The light curve thus peaks
earlier, at a higher luminosity. The ejecta become optically thin
sooner, making the transition to the asymptotic solution of balanced
deposition and radiation at an earlier time, and the peak becomes
narrower than the fiducial model.  The higher-opacity model likewise
peaks later, is fainter, and is considerably broader. Note that this
behavior is the opposite of the Phillips relation; at least for an
opacity which is constant with time, we must look elsewhere for a
fundamental parameter to explain observations. Note also that a factor
of four change in opacity makes only half a magnitude of difference in
the peak magnitude.

Next we show the effect of varying the extent of the energy deposition, but
without varying the mass of \nifsx, the velocity, or the total mass of the
explosion. Such a variation might be the result of hydrodynamically-induced
mixing.  The result is that the models with more centrally-condensed
deposition peak later, but with only very slightly lower peak
magnitude. The width of the peak is somewhat broader with a broader
distribution of deposition as there is a larger range in diffusion times
for the deposited energy to make it to the surface.

We next vary the kinetic energy of the explosions, which is to say the
scale velocity, by a factor of two above and below the fiducial
model. Because a greater expansion velocity leads to a more rapid decline
in column depth, more energetic explosions peak earlier, at higher
luminosities, and decline more rapidly. Thus an increase in explosion
energy has much the same effect as a decrease in opacity. Indeed, the
opacity and the density occur in the thermalized radiation equations only
as the product $\rho\kappa$.  The change in the slope of the energy deposition
following peak is a consequence of the change in column depth to the
$\gamma$-rays.

In the last panel of Figure~\ref{fourparamplot}, the \nifsx mass is varied.
The peak luminosity is seen to follow the \nifsx mass, with a slight change
in shape arising from the varying fraction of the supernova filled with
radioactive material as in the second panel.

Arnett's \citeyear{Arnett82,Arnett96} analytic SNe~Ia light curve
model predicts that the luminosity at bolometric maximum would
precisely equal the instantaneous rate of deposition from \nifsx\ and
\cofsx\ decay, which has provided some interesting constraints, both
on the mass of \nifsx\ produced and on the luminosity at
maximum. ``Arnett's Rule'', as it has been come to be known, is only
approximate however, and related to the assumption that a single
eigenmode describes the shape of the energy density and that the
energy deposition has this same shape.

\begin{figure}[tp]
\postfig{p}{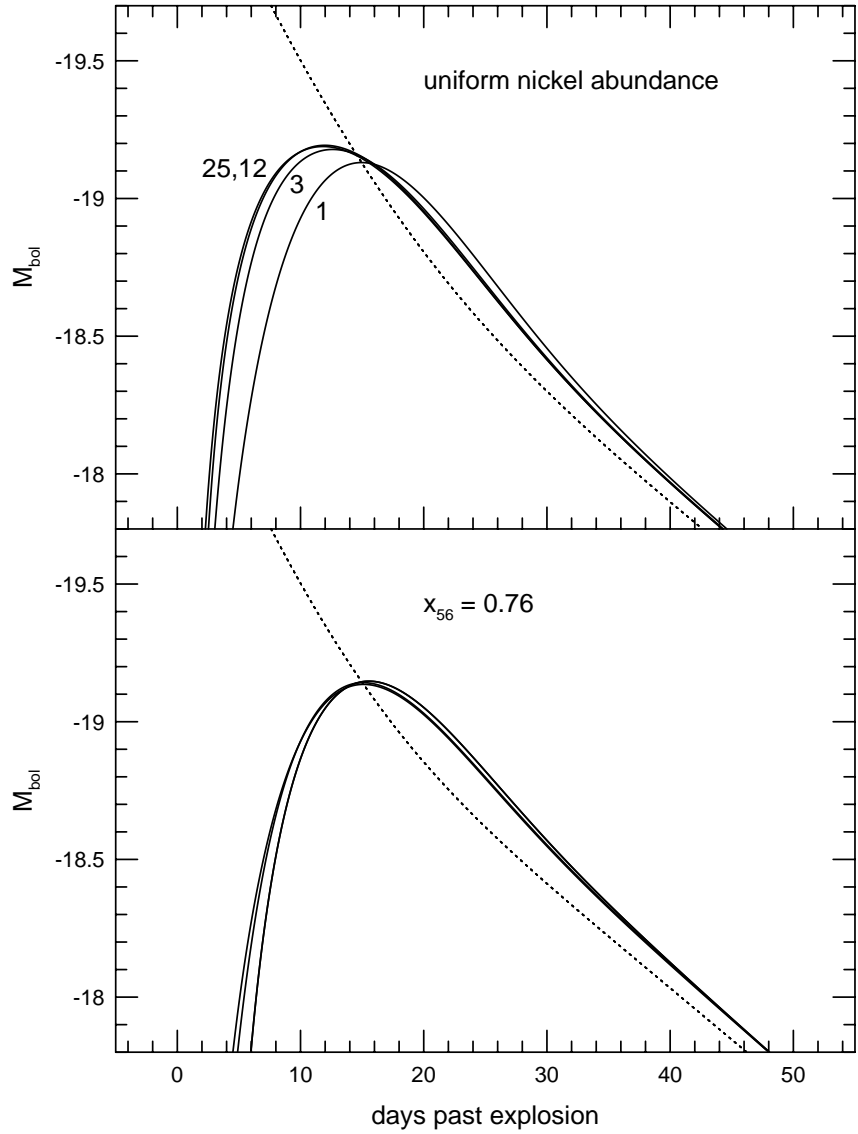}
\caption{The effect of the number of eigenmodes on the calculated
light curve. In the upper panel, energy deposition is taken to be uniform
with radius. In the lower, the deposition extends out to the radius used in
the standard model, x = 0.76.}
\label{twomodeplot}
\end{figure}

Figure (\ref{twomodeplot}) illustrates the result of including a
varying number of eigenmodes in the solution. Arnett's
\citeyear{Arnett82} result is reproduced by taking only the first
mode. The effect of including higher modes is primarily to steepen the
rise to peak and to broaden the width of the peak.  From equation
(\ref{inhomog}) we see that the e-folding time for the power in mode
$n$ to decay is proportional to the eigenvalue, which varies roughly
as the square of the mode number. This is easy to understand
physically. The higher-order modes describe variations of the energy
density on smaller and smaller spatial scales.  The energy variations
at these scales do not have far to go to diffuse out to a smoother
distribution, so the power in these modes declines rapidly.  In the
lower panel of the figure, we have used the same model as in
Figure~\ref{annedd}, while in the upper panel we have made the
energy deposition uniform over the entire star. In both cases the
effect of adding more modes is to steepen the rise to peak. In the
case with the \nifsx ``buried'' well within the ejecta, the energy
from the decay takes some time to diffuse to the surface, by which
time the fundamental mode has most of the power. Thus the time of peak
and the peak magnitude are little affected by the number of modes. For
the uniform-deposition case in the upper panel, however, there is
deposition near the surface which can only be represented adequately
by the inclusion of higher eigenmodes. The energy deposited near the
surface spends less time diffusing and suffers less from adiabatic
decompression. Thus the inclusion of the higher modes shortens the
rise time. The light curve peaks earlier and at a higher
luminosity. Because of this effect, all subsequent light curves in this
work are calculated with a sufficient number of eigenmodes to approach
the exact solution. On the other hand, the figure shows that for
models in which the \nifsx does not extend out beyond, say, 85\% of
the radius, the distribution of radioactivity has little effect on the
light curve. By the time energy has diffused out to the surface,
information about this distribution has been lost.

This tendency toward the fundamental mode near maximum provides a clue to
the expected shape of the peak. For a constant opacity and times long
compared with the scale time, we see from equation~\ref{peakshape} that the
peak of the light curve will be a Gaussian,
\begin{equation}
\phi(t) = \exp\left\{-{{\alpha_1 t^2}\over{2 t_{sc}\tau_d}}\right\}.
\end{equation}
This provides a theoretical justification for the use of a Gaussian as
a fitting function by Vacca \& Leibundgut \citeyear{VaccaL96a} to
determine the risetime and width of observed light curves. For
explosions with significant amounts of \nifsx near the surface, this
approximation will of course be less accurate. In such cases, a larger
number of eigenmodes are necessary to to describe the wider variation
of diffusion times from the sites of deposition to the surface.

The previous figures also show that, except for models with significant
deposition near the surface, the luminosity at peak is identical to the
instantaneous deposition rate (under the assumption of an opacity which is
constant with time, $d\zeta/dt=0$), as first noted by \citeN{Arnett82}.  It
is important to remember that this does {\em not} imply a short diffusion
time at peak. Rather, it results from the fact that peak light is the
watershed which separates times at which the energy deposition rate is
greater than the luminosity from those at which it is less, as noted in the
introduction to this section.

\begin{figure}[tp]
\postfig{l}{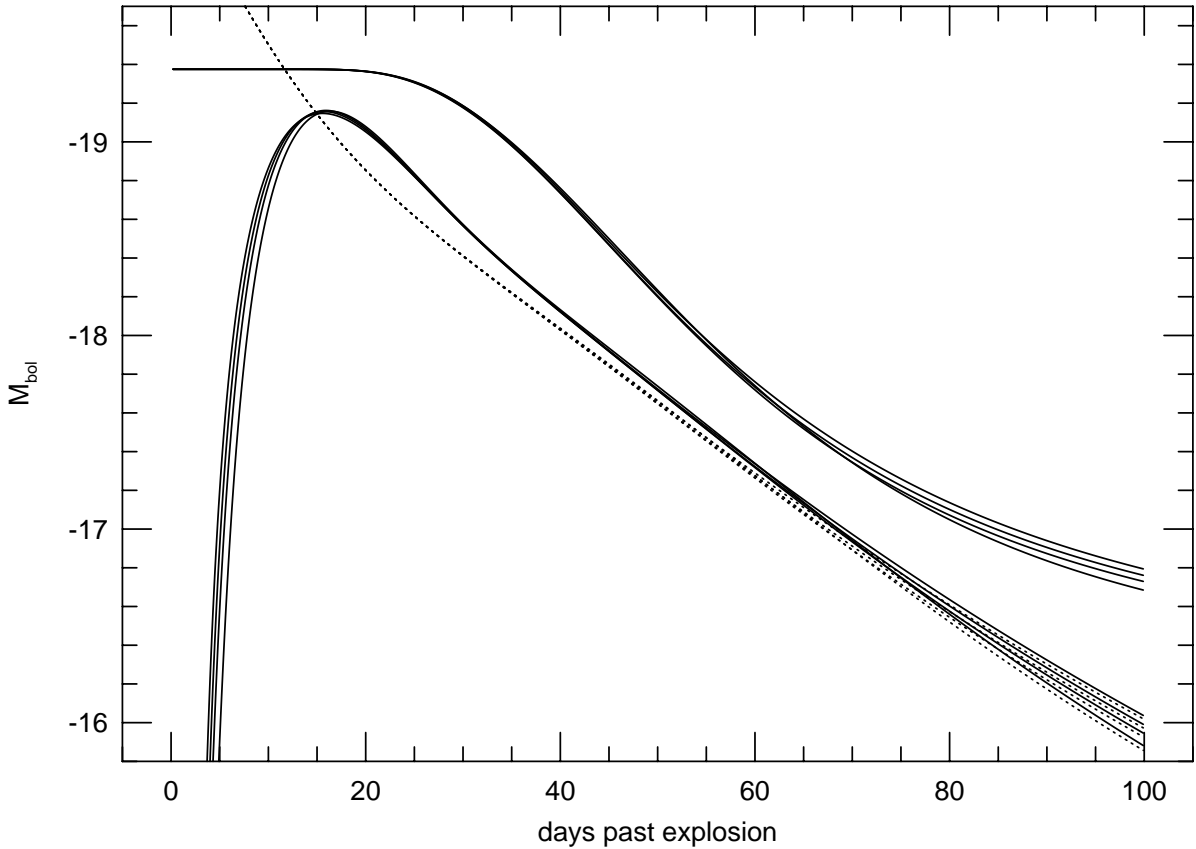}
\caption{A light curve from a model with density structure $\tilde{\rho}(x)
= \exp(-kx)$ plotted along with that of the constant-density fiducial model.
The upper set of curves shows the time dependence of the energy deposition
fraction; for all density laws, the deposition is nearly complete until ten
days or so after peak.
}
\label{dlnrhoplot}
\end{figure}

In Figure \ref{dlnrhoplot} we show the effect of altering the density
structure of the supernova. The density of most SN Ia models is represented
fairly well by an exponential in velocity, $\tilde{\rho}(x) = \exp(-kx)$
with $k\sim 4$, which departs fairly strongly from the constant density
profile we have employed thus far. One can see from the figure that in
spite of the crudeness of the model, a constant density model light
curve is nearly identical to
one which possesses a more realistic density profile. The
light curve is insensitive to the density structure for the same reasons
that it is insensitive to the number of included modes.

None of the parameters we have examined thus far can account for the PR;
varying the explosion energy and opacity lead to a correlation {\em
opposite} in sense to the ``brighter implies broader'' behavior observed,
and the other parameters lead to little variation in light curve shape.  One
way to obtain the PR suggests itself immediately: if the mass of \nifsx is
decreased while the kinetic energy is increased, then a sufficient decrease
in \nifsx mass can offset the increased luminosity of the narrower peak.
The problem with this proposition is that in a \mch explosion, most of the
star must be burned at least to the silicon/calcium group to obtain the
observed velocities. We can lower the \nifsx mass only by increasing the
fraction of the star burned to Si/Ca. Even though approximately 75\% as
much energy is liberated in burning only to Si/Ca as in burning all the way
to \nifsx, it is hard to see how a decrease in \nifsx fraction sufficient
to achieve the desired effect on the light curve can accompany a sufficient
increase in kinetic energy.

The only other way to obtain a ``Phillips Relation'' in an \mch explosion
is to vary the opacity in such a way that an increased \nifsx mass is
accompanied by an increase in opacity enough to offset the increase in
kinetic energy. An increase in \nifsx will in general result in a more
energetic explosion, and hence a narrower peak. If the opacity is increased
sufficiently, however, the peak will be broader nonetheless. An increased
opacity accompanying a higher \nifsx mass might result from a combination
of higher temperatures due to increased deposition and a higher opacity in
iron group elements than in Si/Ca. The results of the following section
call both of these effects seriously into question, however.

\begin{figure}[tp]
\postfig{p}{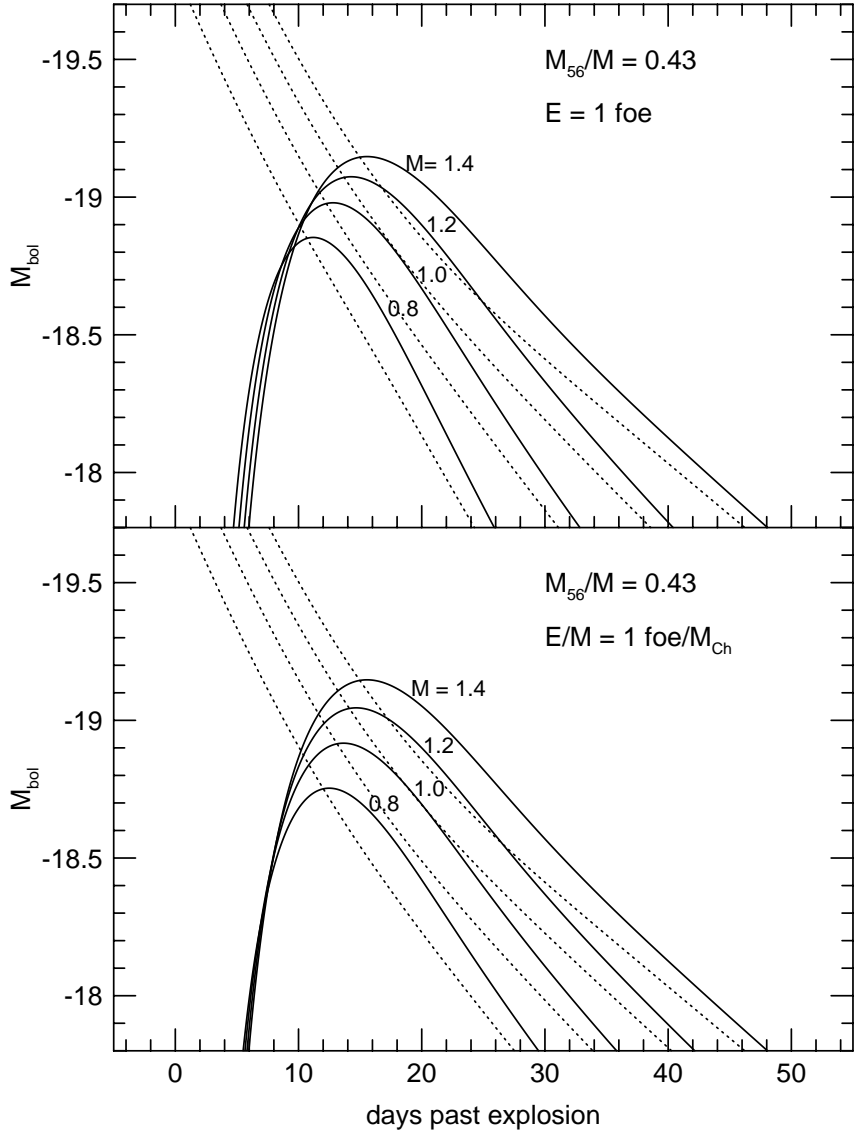}
\caption{Two ways of varying the total mass of the explosion: in the
top panel the explosion energy is constant at 1 foe, and the \nifsx
yield is fixed as a constant fraction of the total mass, for $M=0.8$,
1.0, 1.2 and 1.4~\Msun. In the lower panel $E/M$ is held constant at
$3.6\times10^{17}$~\ergg.}
\label{twomassplot}
\end{figure}

The final parameter in the solution is the total mass of the explosion.
For a constant velocity (specific kinetic energy), changing the total mass
will result in a change in the density and will have a similar effect to
changing the opacity as in figure~\ref{fourparamplot}, leading to a
brighter and narrower peak for lower masses. Lower mass explosions,
however, will naturally produce less \nifsx as the densities attained in
lower mass white dwarfs are smaller.  Changing the density will also affect
the gamma-ray deposition. A decrease in mass will allow energy to escape
more easily in the form of gamma-rays. More importantly, it will allow the
{\em rate} at which the escape increases to be greater, and this more-rapid
fall off in the deposition will also act to oppose the tendency toward
increased luminosity in models with lower column depth.

In Figure~\ref{twomassplot}, the total mass of the ejecta is varied.
Simply as an illustration of the effects, in both panels the \nifsx mass
fraction is kept constant, but in the upper panel the energy of the
explosion is kept constant while in the lower the ratio of explosion energy
to total mass, the specific energy and thus the velocity, is
preserved. Constant-energy explosions might arise, for example, from the
fact that lower-mass white dwarfs have lower densities, leading to an
increasing fraction of the energy arising from incomplete burning to
lighter nuclei. Explosions with constant specific energies would arise when
different mass progenitors nonetheless give similar nucleosynthetic
yields. In both cases, the higher column depth of larger-mass models leads
to later, broader peaks, but the larger adiabatic losses are more than
compensated by the increased mass of \nifsx. In both cases, larger masses
lead to brighter and broader peaks, as observed.

In paper II \cite{PintoE96b} we examine the systematics of
light curves for more realistic models of \sneia both at \mch and below.
For the present, we note that lower-mass explosions would appear to provide
a simple and natural explanation for the physics underlying the PR, with
the total mass of the explosion as the fundamental underlying parameter.

\section{Opacity and Photon Escape}\label{opacity}

\subsection{Opacity Contributors}
If we are to discriminate between models based upon the behavior of their
light curves, it is clear from the above that an accurate understanding and
determination of the opacity is crucial. \citeN{Harkness91},
\citeN{WheelerSH91}, and \citeN{HoflichMK93} have stressed that the
bolometric rise time and peak luminosity depend as sensitively upon the
opacity as on any of the other physical properties characterizing the
explosion: mass, kinetic energy or \nifsx\ mass.  As was shown in the
preceeding section, a factor of two change in opacity has nearly the same
effect on the peak of the light curve as a factor of four change in
explosion energy or a 50\% change in the mass of the ejecta.

In this section we examine the monochromatic opacity in SNe~Ia and the
mechanisms by which energy which has been deposited in the interior
diffuses to the surface. The results of this section will then be applied
to an examination of frequency-averaged mean opacities below.

A simple application of the analytic model presented in section~2 shows
that, for explosion models with $0.7\ \msun <M<1.4\ \msun$ and $0.35\ \msun
<M({}^{56}\hbox{Ni})<1.4\ \msun$, the central temperature near maximum
luminosity is $1.5\times 10^4\ \K\ltsim T \ltsim 2.5\times10^4\ \K$ and the
density is $10^{-14}\ltsim \rho \ltsim 10^{-12}\ \Gcc$. For these
conditions, the continuum opacity at optical wavelengths is dominated by
electron scattering. Central temperatures $T_c>1.5\times10^4\ \K$ mean that
the peak of the Planck function is in the UV ($\lambda_{BB} \ltsim
1900$~\ang) where the opacity is dominated by bound-bound transitions.  The
opacity from a thick forest of lines is greatly increased by velocity shear
Doppler broadening \cite{KarpLCS77}.

\begin{figure}[tp]
\postfig{l}{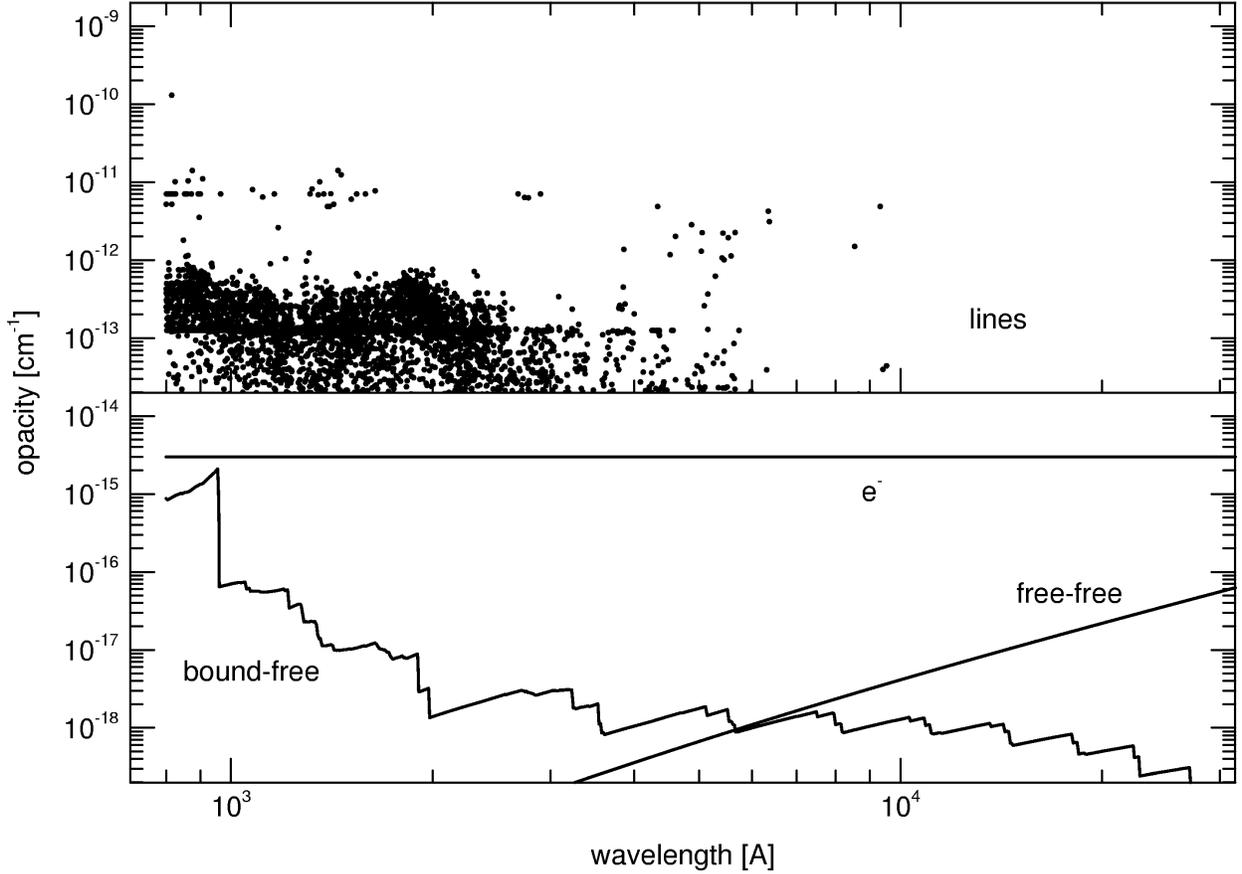}
\caption{Monochromatic opacity sources at maximum light for a
Chandrasekhar mass model of a type Ia supernova from a time-dependent,
multi-frequency, LTE calculation. The physical conditions are:
\protect$\rho=10^{-13}$~\gcc, \protect$T=2.5\times10^4$~K and
\protect$t=14$~days.  The line opacity shown here is the {\sl
expansion} opacity as given by equation~\protect\ref{kappa_exp} (see
text).}
\label{opacplot}
\end{figure}

Figure \ref{opacplot} displays the various sources of opacity for a mixture
of \nifsx (20\%), \cofsx (70\%), and \fefsx (10\%), at a density of
$10^{-13}$~\gcc and temperature of $2.5\times10^4$~K, typical perhaps of
maximum-light in a Chandrasekhar-mass explosion. The excitation and
ionization were computed from the Saha-Boltzmann equation.  The opacity
approximation of Eastman \& Pinto (1993--also see below) was used for the
line opacity, which greatly exceeds that from electron scattering.
Bound-free and bound-bound transitions contribute negligibly to the overall
opacity, but they are important contributors to the coupling between the
radiation field and the thermal energy of the gas.  The opacity is very
strongly concentrated in the UV and falls off steeply toward optical
wavelengths. As was noted by \citeN{MontesW95}, the line opacity between
2000 and 4000\AA\ fall off roughly as $d{\rm ln}\kappa_\lambda /d{\rm
ln}\lambda \sim -10$.  It will be shown that the steepness of this decline
rate toward the optical has important implications for the effective
opacity in \sneia and the way in which energy escapes.

Not only is the opacity from lines greater than that of electron
scattering, it is also fundamentally different in character from a
continuous opacity. In a medium where the opacity varies slowly with
wavelength, photons have an exponential distribution of pathlengths.  Their
progress through an optically thick medium is a random walk with a mean
pathlength given by $(\rho\kappa)^{-1}$.  In a supersonically expanding
medium dominated by line opacity there is a bimodal distribution of
pathlengths. The line opacity is concentrated in a finite number of
isolated resonance regions. Within these regions, where a photon has
Doppler shifted into resonance with a line transition, the mean free path
is very small. Outside these regions, the pathlength is determined either
by the much smaller continuous opacity or by the distance the photon must
travel to have Doppler shifted into resonance with the next transition of
longer wavelength.  For the physical conditions of Figure~\ref{opacplot},
the mean free path of the photon goes from approximately
$5\times10^{14}$~cm in the continuum (due to electron scattering) to less
than $\sim10^{6}$~cm when in a line.  The usual random-walk description of
continuum transport must be modified to take this bimodal distribution
into account.

Within a line, a photon scatters on average $N\sim1/p$ times.
$p$ is the Sobolev escape probability per scattering for escape, which
is accomplished by Doppler shifting out of resonance.  In spite of
this possibly large number of scatterings needed for escape, a photon
spends only a small fraction of its flight time in resonance with
lines.  The effect is quite different from that of a similar
observer-frame optical depth arising from a continuous opacity.

Because of the very supersonic expansion of the supernova's ejecta, we can
make use of Sobolev theory to describe the path of a photon, following the
discussion of \citeN{EastmanP93}. The Sobolev optical depth of a line
transition with Einstein coefficient $B_{lu}$ is (ignoring stimulated
emission)
\begin{equation}
\tau_s = {{h}\over{4\pi}}{{n_lB_{lu}}\over{|\partial\beta/\partial l}}
\end{equation}
where $n_l$ is the lower level number density and $\partial\beta/\partial
l\sim1/ct$ is the velocity gradient over the speed of light. The probability of
escape from the line transition is
\begin{equation}
 p  = {{1 - e^{-\tau_s}}\over{\tau_s}}.
\end{equation}
\cite{Castor70}. Consider a photon which is emitted in resonance at
frequency displacement $x$ and subsequently re-absorbed at displacement
$x^\prime<x$, having travelled a distance
$(x-x^\prime)\Delta\nu_D/(\partial\beta/\partial l)$, where $\Delta\nu_D$
is the thermal Doppler width. Assume for the moment that the line has a
negligible photon destruction probability. We will address thermal
destruction below.  The optical depth between emission and absorption is
\begin{equation}
\tau(x,x^\prime) = \tau_s \int_{x^\prime}^x \phi(t) dt,
\end{equation}
where $\phi(x)$ is the normalized line absorption profile. The probability
that a photon emitted at $x$ will travel a
distance $(x-x^\prime)ct$ to be re-absorbed at $x^\prime$ is
\begin{equation}
\tau_s \phi(x) \phi(x^\prime) exp\left\{-\tau(x,x^\prime)\right\}.
\end{equation}
Assuming complete redistribution, the average value of $x-x^\prime$ is then
\begin{equation}
<x-x^\prime> = 
{{\tau_s
\int_{-\infty}^\infty \phi(x) \int_{-\infty}^x (x-x^\prime) \phi(x^\prime)
exp\left\{-\tau(x,x^\prime)\right\} dx^\prime dx
}
\over
{ 1 - \left(1 - e^{-\tau_s}\right)/\tau_s }}.
\end{equation}
Here, the denominator is just the total probability of reabsorption,
\hbox{$1-p$}. For a Doppler line profile, this can be approximated as
\hbox{$<\!x-x^\prime\!>\sim 0.8/(1+\tau_s/5)$} \cite{EastmanP93}.  The
typical distance travelled between scatterings in the transition is then
\begin{equation}
\delta r = <\!x-x^\prime\!> 
{{\Delta\nu_D}\over{(\partial\beta/\partial l)}}.
\end{equation}
For homologous expansion $(\partial\beta/\partial l) = 1/ct$. 
On average, the photon will scatter $N \sim 1/ p$ times, and
the total distance covered while trapped in the line resonance will be
\begin{equation}
\delta r_L = \left(1/ p - 1\right) \delta r.
\end{equation}
In a very
optically thick line, the photon will thus travel a distance
equal to $4 v_{th}/v_{exp}$ times the radius of the ejecta while trapped
within a line. Since $v_{th}/v_{exp}<10^{-2}$ at all times, the
photon travels a negligible distance, and hence spends a negligible time
in resonance with any one line. 

What this means is that each time an optically thick line absorbs a
photon, the photon is almost instantaneously re-emitted in a random
direction. From the point of view of the diffusion of radiation
through the ejecta, each line resonance interaction acts like a single
scattering event, independent of the optical depth of the transition!
The distribution of mean free paths will thus be determined by the
continuum opacity and the distribution of lines in energy.  In a
medium with little continuous opacity the mean free path of a photon is
the average distance a photon travels between resonances. The
effective mean free path has little to do with {\em any}
conventionally-defined monochromatic opacity.

For the purpose of estimating the diffusion time, the effective total
``optical depth'' of the supernova for a photon emitted from the
center of the remnant at frequency $\nu$ is the {\em number} of lines
interactions a photon undergoes in Doppler shifting from $\nu$ to
$(1-v_{exp}/c)\nu$. This optical depth may be written as
\begin{equation}
\tau(\nu) = \sum_{\{k|\nu\ge\nu_k\ge\nu(1-v_{exp}/c)\}}
\left(1-\exp(-\tau_k)\right)
\label{taueff}
\end{equation}
The sum is over all lines with Sobolev optical depths $\tau_k$ and
transition frequencies, $\nu_k$, lying in the interval
$\nu\ge\nu_k\ge\nu(1-v_{exp}/c)$.

The evolution of this optical depth with time depends on whether the lines
are optically thick or optically thin.  In the limit that all lines are
optically thick, $\tau_k>>1$, $\tau(\nu)$ is just the number of lines in
the range $(\nu(1-v_{exp}/c),\nu)$ and, barring significant changes in
excitation conditions, is constant with time. In the other extreme, where
all lines are optically thin, $\tau_k<<1$ and $\tau(\nu)=\sum_l
\tau_l$. Since $\tau_l\propto t^{-2}$ (again, barring changes in excitation
conditions), $\tau(\nu)\propto t^{-2}$, and behaves like a continuum
optical depth which is proportional to the ejecta column density.

We can derive an effective monochromatic opacity coefficient by setting
$\rho \kappa(\nu) R_{max} =\tau(\nu)$. This would correspond to a global
average over the frequency range $(\nu(1-v_{exp}/c),\nu)$.  A more local
quantity is obtained by reducing this range to $(\nu,\nu+\Delta\nu)$, where
$\Delta\nu\sim\nu\Delta r \partial\beta/\partial r$, giving
\begin{equation}
\kappa(\nu) = {\nu\over \rho \Delta\nu}{\partial\beta\over\partial r}
\sum_{\{k|\nu\le\nu_k\le\nu+\Delta\nu)\}}
\left(1-\exp(-\tau_k)\right).
\label{kappa_exp}
\end{equation}
This is the expansion opacity given in \cite{EastmanP93}. It has the
advantage of being a purely {\em local} quantity. The expansion opacity
formulation of \citeN{KarpLCS77} and its descendents, on the other hand,
always averages over a mean free path. This distance can easily become
larger than the distance over which the material properties change or even
than the supernova itself.

\begin{figure}[tp]
\postfig{l}{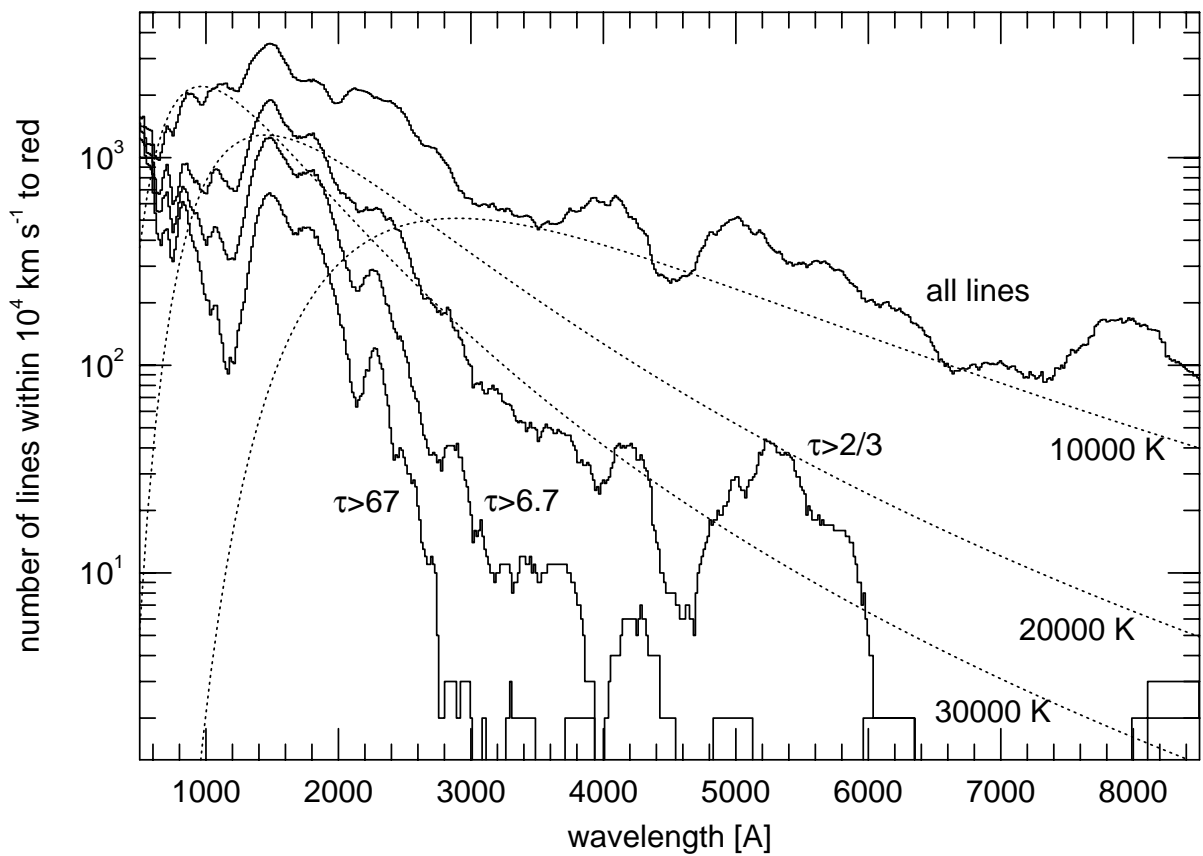}
\caption{The density of lines in energy space at maximum light (18
days). The histograms are the number of lines in $10^4$~\kms\ to the
red of a given wavelength, binned in energy. The heavy line includes all lines
with an average (over volume) Sobolev optical depth greater than unity.
The two lower histograms show the same quantity, but include lines with
average Sobolev optical depth greater than 6.7 and 67. Roughly speaking,
the optical depth one curve will decline to resemble the optical depth 6.7
curve at 31 days.  Data are taken from the same calculation as for Figure
(\protect\ref{opacplot}). The dotted lines are schematic flux distributions for
blackbody radiation fields of one, two, and three $\times10^4$~K for
comparison.}
\label{linedensity}
\end{figure}

In terms of the effective total optical depth, $\tau(\nu)$, the
diffusion time can be written $t_d \approx \tau(\nu) R(t)/c$. By
setting $t_d=t$, substituting $R(t) = v_{exp} t$ and $v_{exp}\sim 10^4$~\kms, 
one finds that $\tau(\nu)\sim c/v_{exp} \sim 30$. In the UV,
$\tau(\nu)>>30$, and since the Sobolev optical depth of many of these
lines is $>>1$, $\tau(\nu)$ will remain approximately constant with
time. The fact that there is a peak in the light curve, which occurs
when $t_d=t$, means that the {\it flux mean} optical depth must
drop below $\sim30$.

In Figure \ref{linedensity} we show $D(\nu)=$ the density of lines per
$10^4$~km~s$^{-1}$, for the same conditions as were used for Figure
\ref{opacplot}, at 18 days past explosion, with lines taken from the Kurucz
line list \cite{Kurucz91}. The uppermost curve is the spectral density of
all lines included in the calculation. Below that is all lines with
$\tau_s>2/3$, and further down are all lines with $\tau_s>6.7$ and 67. If
we exclude lines for which $\tau_s\ltsim1$, then $D(\nu) \approx \tau(\nu)$.
Superimposed upon these curves for reference are three blackbody
distributions (the vertical scale is arbitrary). For temperatures above
$10^4$~K, most photons see a value of $\tau(\nu)$ greatly in excess of the
critical value derived above and remain trapped with an ever-increasing
diffusion time. Since individual line Sobolev optical depths decline as
$1/t^2$, it will not be until ~50 days when the $2\times10^4$~K Planck mean
of $\tau(\nu)$ falls below 30. What, then, accounts for the fact that the
light curve peaks at 18~days and not 50~days? One possibility may be, at
least in part, that as photons random walk their way out, they are Doppler
shifted to longer wavelengths where the effective optical depth is much
smaller.

If photons must scatter on order $\tau(\nu)^2$ times to escape, and the
mean free path is $R/\tau(\nu)$, we can ask what the accumulated Doppler
shift might be, following this path. Since, for homologous expansion $dv/dr
= v_{max}/R=1/t$, the total Doppler shift is $\tau(\nu) \nu v_{max}/c \sim
\nu$, i.e. of order of the entire energy of the photon!  This implies that,
in the absence of photon destruction mechanisms (electron collisional
de-excitation, branching), a photon emitted in the interior will scatter off
lines until it has accumulated sufficient redshift to put it at a frequency
where $\tau(\nu)$ is small enough to permit escape.

\subsection{Photon Collisional Destruction and the Thermalization
Length}\label{PhotCollSec}

The discussion of line opacity has so far assumed that lines are coherent
scatterers, with no account taken of photon ``destruction''. By this we
mean any alternative channel for de-populating the upper state other than
emission and escape of a photon in the original transition. The two most
important mechanisms for this are branching and collisional depopulation by
thermal electron collisions.

\begin{figure}[tp]
\postfig{l}{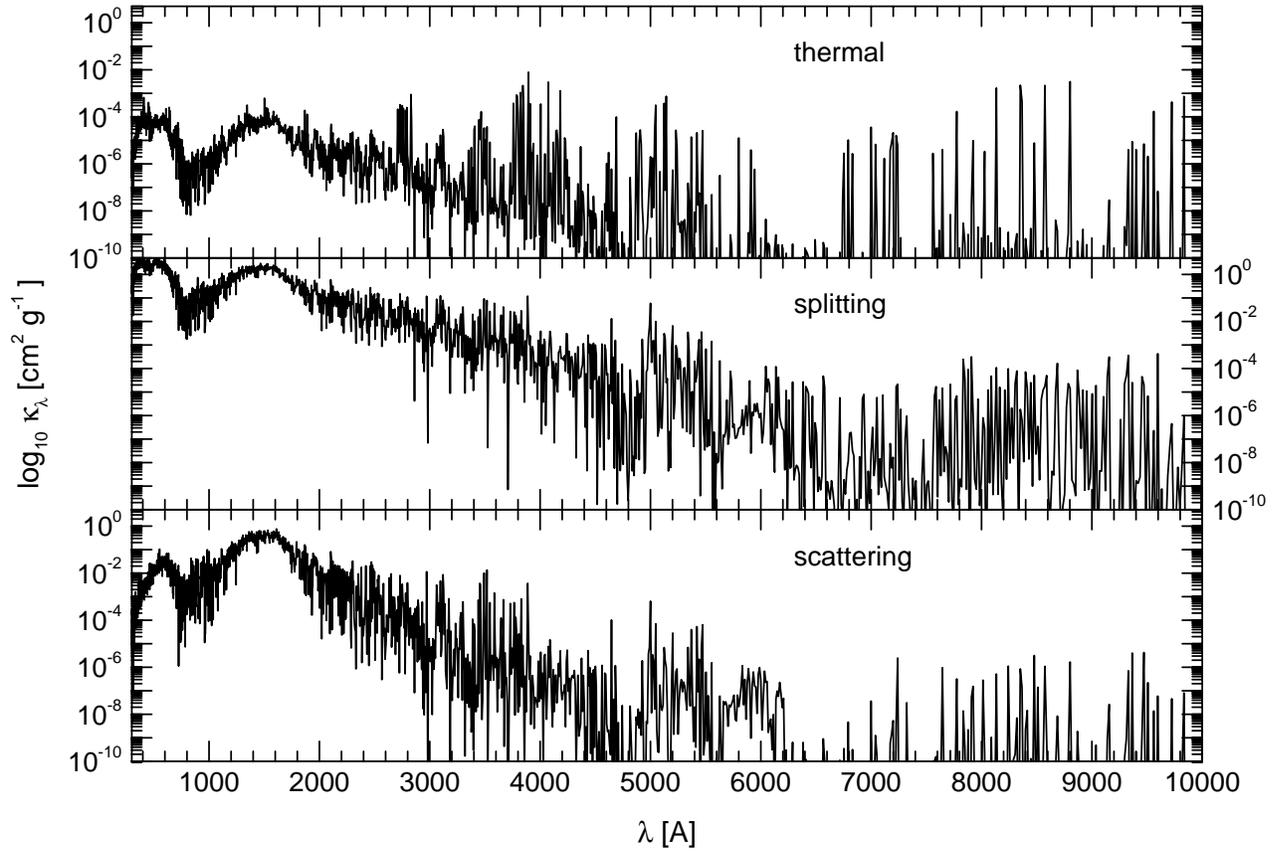}
\caption{ The opacity of a gas at the same conditions as in Figure
\protect\ref{opacplot}, decomposed into thermal destruction (top:
equation~\protect\ref{kappa_thm}), photon splitting (middle:
equation~\protect\ref{kappa_spl}), and coherent scattering (bottom:
equation~\protect\ref{kappa_scat}).  LTE level populations were
employed.}
\label{scat_therm_split}
\end{figure}

We first examine the efficiency for collisional destruction, that upon
being excited to the upper level of the transition, the absorbing atom
is collisionally de-populated and the photon's energy is added to (or
subtracted from) the thermal kinetic energy of the gas. The
collisional destruction probability per line interaction to upper
level $u$ ({\em not} per single scattering) can be written
\begin{equation}
\epsilon_u = {{n_e \sum_l C_{ul}}\over{n_e \sum_l C_{ul} + \sum_l
 p_{ul}A_{ul}}}
\label{epsilonu}
\end{equation}
where, for $E_u>E_l$
\begin{equation}
n_e C_{ul} = {8.629\times10^{-6} \Omega_{ul}(T)n_e\over g_u \sqrt{T}}
\end{equation}
(cf. \citeNP{Osterbrock89}) is the rate per atom of collisions from state
$u$, with statistical weight $g_u$, and $ p_{ul}A_{ul}$ is the effective
radiative de-excitation rate.  To investigate the effect of electron
collisions on the effective line opacity one can use equation~\ref{kappa_exp},
with each line $k$ weighted by the probability for thermalization:
\begin{equation}
\kappa_{thm}(\nu) = {\nu\over \rho \Delta\nu}{\partial\beta\over\partial r}
\sum_{\{k|\nu\le\nu_k\le\nu+\Delta\nu\}}
\left(1-\exp(-\tau_k)\right)\times \epsilon_k
\label{kappa_thm}
\end{equation}
where $\epsilon_k$ is the thermalization probability for line $k$
(equation~\ref{epsilonu}) and $\Delta\nu$ is the frequency bin size.
The top panel of Figure~\ref{scat_therm_split} shows $\kappa_{thm}$
for the same line list and conditions as in Figure~\ref{opacplot}. Van
Regemorter's formula \cite{VanRegemorter62} was used to calculate the
collision rates, $\Omega_{ul}$ but in no case was the resulting
collision strength allowed to be less than unity. For this particular
example, $\kappa_{thm}$ peaks at a value of $\sim10^{-4}\ \hbox{cm}^2$
near 1500~\ang. To put this in context we must consider the question
of what value of $\kappa_{thm}$ is sufficient to bring the gas and
radiation field into thermal equilibrium. This will occur when
$\tau_{thm}\equiv\rho R_{max}\sqrt{\kappa\kappa_{thm}}\gtsim 1$, where
$\kappa$ is the total opacity. For a 1.4~\Msun uniform density sphere
expanding at $10^9$ \kms, the column density at 18~days is $\rho
R_{max}\sim276$~\gcmsq. At 1500~\ang,
$\kappa_{thm}\sim10^{-4}$~\cmsqrgm, while the total opacity is
$\kappa\sim1$~\cmsqrgm, so $\tau_{thm}\sim 2.8$ ---barely adequate to
thermalize the radiation field to the local gas temperature.  Longward
of 1500~\ang\ the situation is somewhat different.  At 3000~\ang, for
instance, $\kappa\sim10^{-3.5}$ and $\kappa_{thm}\sim10^{-6}$, so
$\tau_{thm}\sim5\times10^{-3}$---much less than 1 and entirely
insufficient for thermalization.

While these numbers should be taken as no more than order of magnitude
estimates, they are accurate enough for us to conclude that near maximum
light, the electron density in Type~Ia supernova ejecta is too low for
collisional destruction to bring about thermalization between gas and
the radiation field, at least at wavelengths
$\lambda\gtsim2000$~\ang. We expect therefore that the radiation field
at longer wavelengths may be significantly different from a Planck
function at the local gas temperature. This is a very important point.
It means that there is no depth in the supernova to apply the usual,
equilibrium radiative diffusion inner boundary condition wherein it is
assumed that $J(\nu) = B(\nu,T_{gas})$.  We show below that line
scattering results in a pseudo-continuous spectrum, however this must
not be confused with a continuum that arises from thermalization
mediated by electron collisions.

\subsection{Photon Splitting and Enhanced Escape}

Another possible fate for a photon trapped in a line resonance is that
it will decay from the upper level, not in the transition it was
absorbed in, but to some other state of lower energy. The probability
that a photon trapped in a resonance with upper state $u$ will be
escape the resonance region in via downward transition $l$, is given
by the branching ratio
\begin{equation}
b_{ul} = {{ p_l A_l}
\over{n_e \sum_j C_j + \sum_k p_k  A_k}}
\label{bul}
\end{equation}
where the sum $k$ is over all lines with the same upper level $u$,
including line $l$. While UV lines tend to have the highest Einstein $A$
values (for dipole permitted transitions $A\propto\lambda^{-2}$) they also
tend to arise from levels nearest the ground level, to have higher optical
depths and, therefore, smaller escape probabilities. This may lead to a
larger probability of decay into another, longer wavelength transition, and
further cascade into yet other transitions. Let us call this process
``photon splitting'', as the effect is to split a photon's energy up into a
series of longer-wavelength photons. This process is well observed in
nebul\ae, where each photon of higher-energy transitions in the Lyman
series of hydrogen is ``degraded'' to emerge from the lowest-energy
transitions in lower-energy series.  The energy in Lyman-$\gamma$ photons,
for example, is ``split'' into Lyman-$\alpha$, Balmer-$\alpha$, and
Bracket-$\alpha$ photons.  While the inverse process, absorption from a
high-lying state and decay into a higher-energy transition, can and does
occur, it must do so less frequently by obvious thermodynamic
considerations (Rossleand's Theorem of Cycles---cf. \citeNP{Mihalas78}).
\citeN{Pinto88} noted the importance of splitting to spectrum formation in
\sneia in the nebular phase. \citeN{LiM96} showed that it is an important
effect in SN~1987A at late times as well.

\landscale{1.0}
\begin{figure}[tp]
\postfig{p}{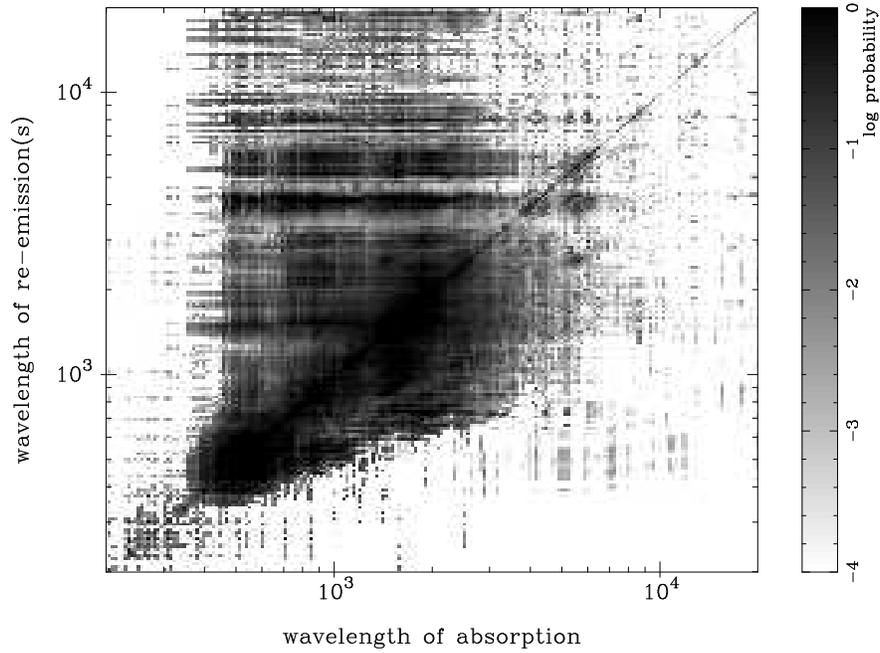}
\caption{The cascade matrix for a photon absorbed into a transition with
wavelength given by the abscissa and emitted into (possibly many)
wavelengths given by the ordinate. The intensity represents the the
probability of branching multiplied by the probability of being trapped in
the absorbing transition. Thus, absorptions at low optical depth, which may
nonetheless lead most likely to multiple splittings, are suppressed.
The line list and physical conditions are again the same as for
Figure \protect\ref{opacplot}.}
\label{cascadeplot}
\end{figure}
\defaultscale

In Figure \ref{cascadeplot} we show the relative probability that a
photon's energy, absorbed into a transition at a given wavelength, is
re-emitted at one or more other wavelengths, weighted by the probability of
absorption into the initial transition.  It is much like the more familiar
recombination cascade matrix, but instead of a collisional process pumping
the high energy states, in this case it is line absorption. There is a
considerable tendency for energy absorbed in the UV to come out in the
optical and infrared. This tendency is enhanced by the radiative transfer
through the ejecta, as a photon emitted in the UV is overwhelmingly likely
to be re-absorbed in another thick transition and given another opportunity
to split, while those photons emitted at longer wavelengths are more likely
to escape.

In addition, the rate of electron collisions coupling states of similar
energies is much larger than those which remove a substantial fraction of
the photon's energy to the gas.  These collisions between states of similar
energies have the effect of opening up an even larger number of subordinate
transitions, enhancing the rate of ``splitting''.

As was done for electron collisional destruction, we can define an
effective line opacity for splitting as
\begin{equation}
\kappa_{spl}(\nu) = {\nu\over \rho \Delta\nu}{\partial\beta\over\partial r}
\sum_{\{k|\nu\le\nu_k\le\nu+\Delta\nu)\}}
\left(1-\exp(-\tau_k)\right) \times \sum_{l\ne k} b_{u(k)l}
\label{kappa_spl}
\end{equation}
where $b_{u(k)l}$ is as given by equation~\ref{bul} and $u(k)=$ the
upper level of transition $k$. The sum is over all downward
transitions except $l$.  We can similarly
define an effective scattering opacity as the line opacity multiplied
by the probability that a photon absorbed into a transition ultimately
escapes as a single photon at the same energy:
\begin{equation}
\kappa_{scat}(\nu) = {\nu\over \rho \Delta\nu}{\partial\beta\over\partial r}
\sum_{\{k|\nu\le\nu_k\le\nu+\Delta\nu)\}}
\left(1-\exp(-\tau_k)\right) b_{u(k)k}
\label{kappa_scat}
\end{equation}
Examples of $\kappa_{spl}(\nu)$ and $\kappa_{scat}(\nu)$ are shown in 
the middle and bottom panels of Figure~\ref{scat_therm_split}, respectively.

With the exception of a few optical wavelengths where $\kappa_{thm}$
dominates, the bulk of the line opacity may be described as primarily
``splitting'' and ``scattering''. This then provides another path for
energy to escape from the supernova. Even deep within the ejecta where
the UV optical depth is quite great, there is a significant ``leak''
of energy {\em downward in frequency} to energies where the optical
depth is much lower. Splitting is a much more efficient mechanism for
downgrading photons in energy than the Doppler shift accumulated
through repeated scatterings described in the previous section.

\begin{figure}[tp]
\postfig{l}{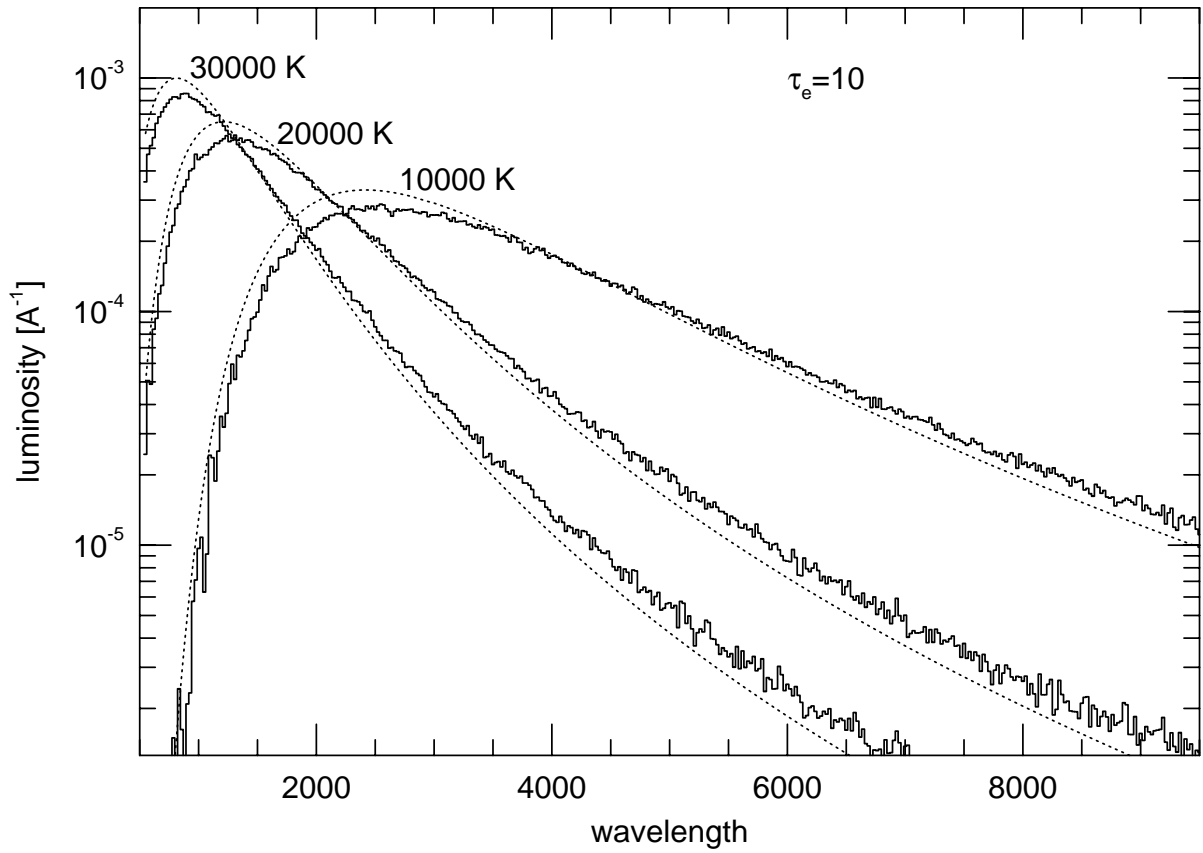}
\caption{ Photons with a Planck distribution in energy and at three
different temperatures were emitted uniformly throughout the volume of
a constant density sphere and followed by a Monte Carlo simulation
until emergence.  The model has an outer velocity of $10^9$~\kms and
total, pure-scattering optical depth of $\tau_e = 10$.}
\label{monte_e}
\end{figure}

It is instructive to examine the results of a few simple, schematic
models. In these, the supernova is taken to be a constant-density sphere
expanding homologously with an outer velocity of $10^4$~\kms. Photons are
emitted uniformly throughout the volume of this sphere with a Planckian
energy distribution. While emitting the photons at the center of the sphere
(and hence at a larger average optical depth) would have provided more
extreme demonstrations of the scattering physics, uniform emission is
closer to the case in a real supernova. The fate of a large number of
emitted photons is followed with a simple Monte Carlo procedure.

The first calculation illustrates the progressive redshift of trapped
radiation which results from multiple scatterings.  The opacity is due
only to scattering coherent in the co-moving frame. The total optical
depth of the ``supernova'' is 10. This corresponds to the electron opacity
of a Chandrasekhar mass with an outer velocity of $10^4$~\kms at 15
days past explosion, ionized on average to Co~V. In this calculation a
typical photon looses 8\% of its energy before escaping.

\begin{figure}[tp]
\postfig{l}{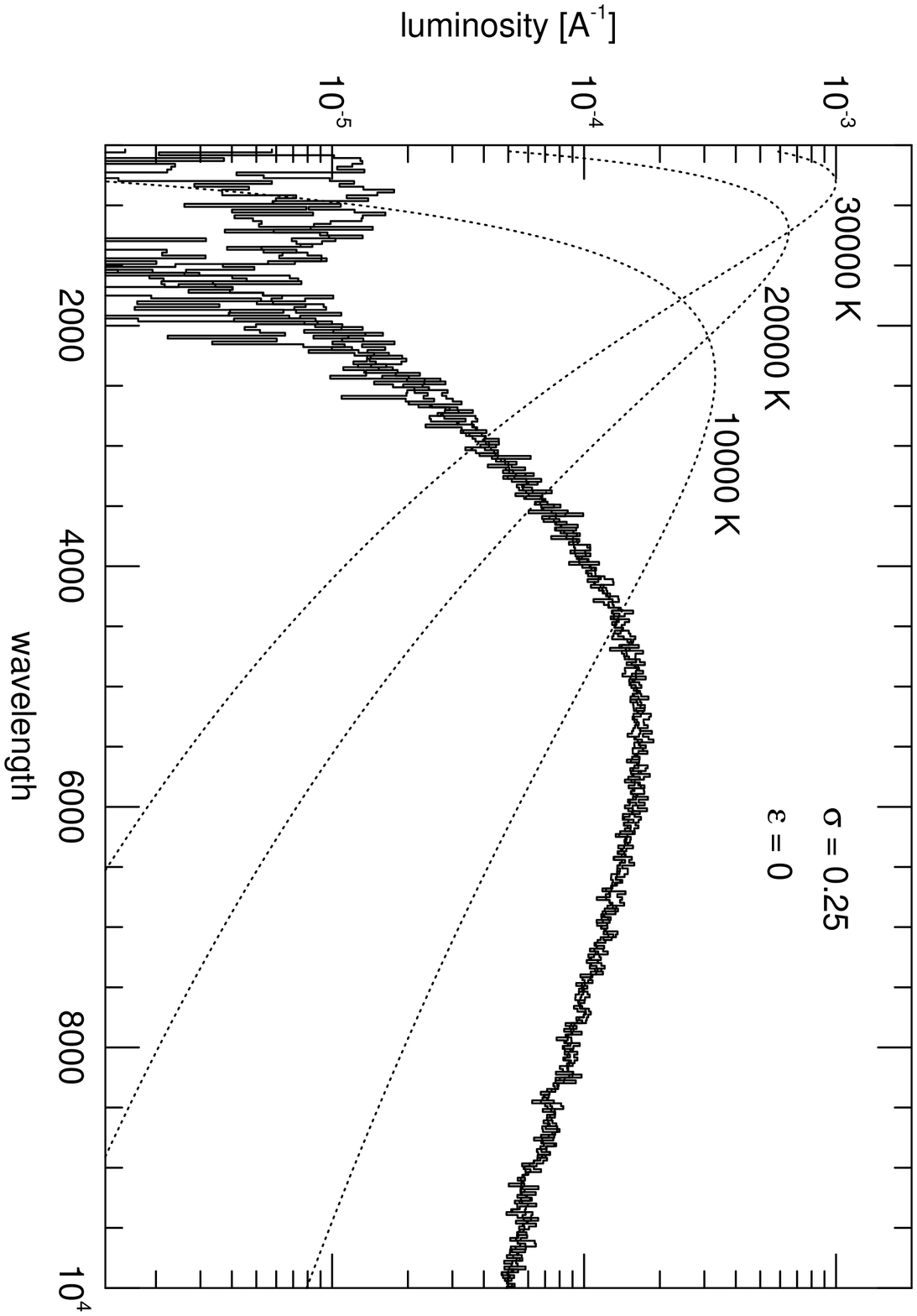}
\caption{An initially Planckian radiation field at three different
temperatures (dotted lines) were transported through the line opacity
described in the text. The total injected energy is the same in each case,
and the emergent spectra (histograms) are virtually identical.}
\label{monte_temp}
\end{figure}

In the next calculations we have used a picket-fence opacity with a line
density $D(\lambda)$ which is a smooth power-law approximation to that
shown in Figure \ref{linedensity}:
\begin{equation}
D(\lambda) = \left\{ \begin{array}{ll}
400 & \lambda < 1500 \\
10^3 (\lambda/1500)^{-4.5} & \lambda > 1500
\end{array}
\right.
\end{equation}
All of the lines have a Sobolev optical depth of 10 (consequently,
$\tau(\lambda)=D(\lambda)$), although the result is unchanged with any
value of the line optical depth greater than unity. Upon absorbing a
photon, a line has a given probability $\sigma$ of splitting into a pair of
photons with longer wavelengths. Energy is conserved by requiring that the
combined photon energies equals that of the absorbed photon. This is a far
more random splitting than that imposed by a real set of atomic data where
each line has a fixed set of branching ratios into a finite (though
typically quite large) number of wavelengths. It nonetheless captures the
spirit of the photon splitting process described above and is
computationally expedient. In these calculations we have ignored electron
collisional destruction.

In Figure \ref{monte_temp} we have taken the splitting probability to
be $\sigma=0.25$ and computed three spectra corresponding to emission
temperatures of $1, 2,$ and $3\times 10^4$~K, but in each case with
the same total emission rate.  The result of this calculation is
remarkably different in character from the coherent scattering case.
One can see from the figure that as long as the peak wavelength of the
blackbody emission spectrum, $\lambda_{Wien}$, lies at a wavelength
characterized by $\tau(\lambda_{Wien})>>1$, the {\em shape} of the
emitted flux is identical in all cases; any memory of the shape of the
thermal emission spectrum has been lost.

\begin{figure}[tp]
\postfig{l}{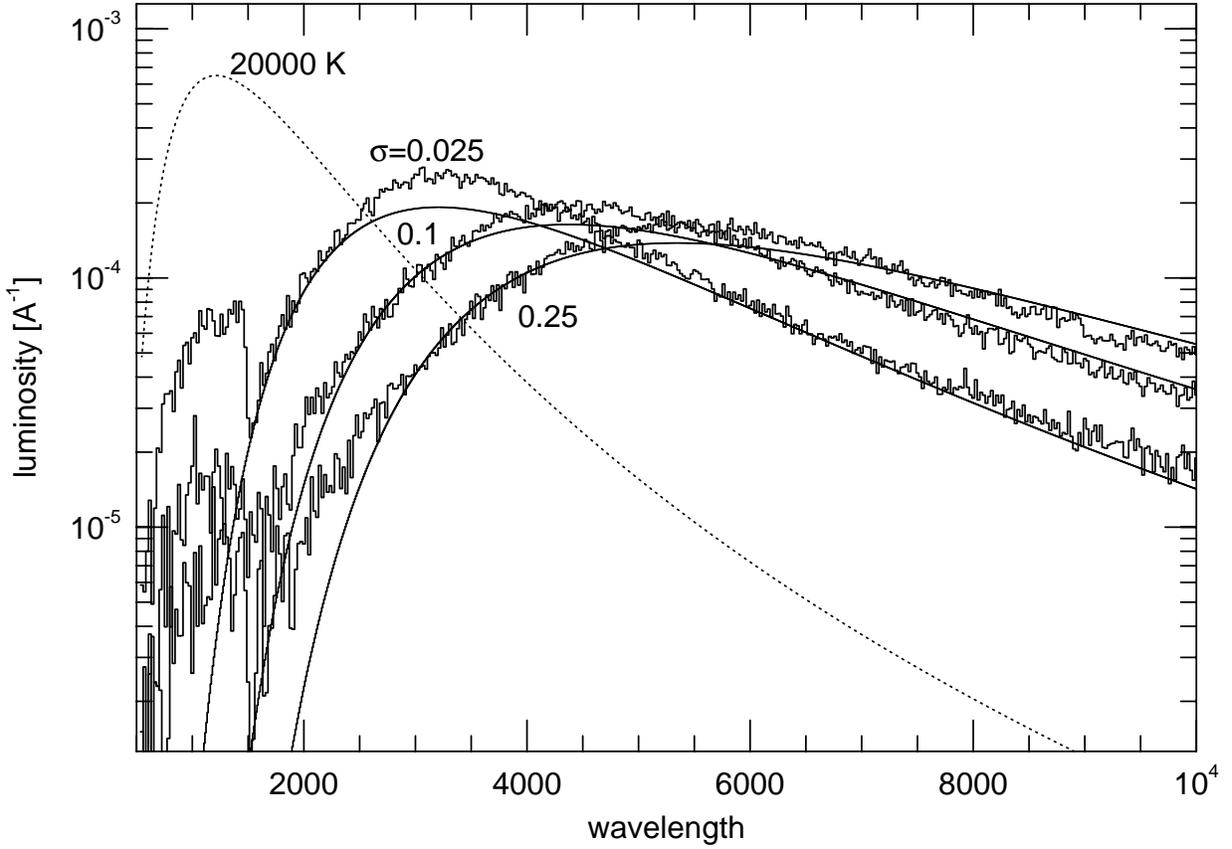}
\caption{Photons from the blackbody shown by the dotted line were scattered
through the line opacity described in the text. While the shape of the
emergent spectrum (histograms) is sensitive to the splitting probability
$\sigma$, it is independent of the input blackbody's temperature.  The
solid lines following the histograms are blackbodies at 4500, 5500, and
7500~K.}
\label{monte_sigma}
\end{figure}

Figure \ref{monte_sigma} shows the effect on the emergent spectrum of
varying the splitting probability. We have chosen three values -- 0.25, 0.1,
and 0.025 -- and a single emission spectrum at $2\times10^4$~K. In all three
cases the emergent spectrum is completely insensitive to the temperature of
the emission spectrum (as established by calculations similar to those of
Figure \ref{monte_temp}). However, it is quite sensitive to the splitting
probability $\sigma$. A remarkable feature of these calculations is that
the emergent spectrum comes very close in shape to a Planck spectrum. The
solid lines in the figure which follow the emergent spectra are blackbodies
at temperatures of 4500, 5500, and 7500~K. In detail, the emergent flux has
a Rayleigh-Jeans tail and a Wein cutoff, and is just slightly more peaked
than an actual blackbody; they have a small, but non-zero, chemical
potential. The peak wavelength, $\lambda_{peak}$, is determined by the
condition $\sigma\tau^2(\lambda_{peak})\sim1$. Any departure in the
spectral shape from an exact Planck function would most likely be impossible
to discern in spectra of real supernovae as the pseudo-continuum produced
by multiple line splittings has superimposed upon it a number of strong
lines as well as structure from deviations in $D(\lambda)$ from the smooth
power-law employed for these examples.

The fact that one can use the basic physical picture these examples outline
to produce a spectrum which is nearly a Planck function in shape but which
has nothing to do with the ``thermal emission'' leads one to suspect that
observed color temperatures may have less to do with close thermal coupling
between gas and radiation field and more to do the distribution of lines
and branching ratios in the complex atomic physics of the iron group.

The physical process underlying the production of these
pseudo-blackbodies is simple: the energy in short wavelength photons
is continuously redistributed to longer wavelengths until an
equilibrium is reached between the injection of energy and its loss
from the system.  The similarity to equilibrium thermodynamics
suggests an interesting possibility. If the redistribution in energy
from multiple splittings is sufficiently large and random to reach a
thermodynamic limit, perhaps a calculation which assumes this
thermodynamic limit might not be too much in error.  The occupation
numbers of the atomic states in the scattering system might not be too
far from their LTE values. In this case, treating the lines as pure
absorbers and using LTE level populations might not be a bad
approximation.  The approach to this thermodynamic limit is
fundamentally different, however.  In a gas, the distribution of
particles in phase space approaches a Maxwellian distribution because
of the essentially infinite number of ways collisions can redistribute
momentum. In the usual LTE radiation transport picture, either a large
bound-free continuum optical depth or the dominance of collisional
rates over radiative ones ensures that the thermodynamic equilibrium
established by gas particles is strongly coupled to the radiation
field, giving it an equilibrium distribution as well. This is in spite
of the relatively restricted possibilities for energy redistribution
in the photon gas through radiative processes. In the present case,
the extreme complexity of the radiative processes allows the photon
gas to reach something of a thermodynamic equilibrium on its own.
This equilibrium will then drive the level populations to LTE values
through radiative processes.  We may expect, then, that the electron
gas will be driven toward a similar equilibrium state through its
(albeit weak) coupling to the photon gas.  In a less schematic
calculation, of course, the emergent spectrum would reflect the gaps
and bumps of Figure \ref{linedensity}, and the result would be a line
spectrum similar in character to that emitted by a supernova.  The
important point is that the shape of the emergent spectrum reflects
the variation in $D$ and not the underlying radiation temperature.

Several sources of uncertainty exist regarding the line-blanketing
opacity. One involves the atomic data. The line list most commonly employed
is that of \citeN{Kurucz91}. Because the number of lines which contribute
to the total opacity is so large, numbering in the hundreds of thousands or
more, there is no way at present to know either how accurate nor how
complete the list is with respect to the weaker lines. The bulk of the
opacity comes from the cumulative effect \cite{Harkness91} of many weak
lines of iron group elements and the lack of any information about
completeness in the data set must be regarded as an unknown source of
uncertainty in the calculations.

Another uncertainty in the line-blanketing opacity arises from
approximations in its numerical representation. It has been noted by
several authors (e.g. \citeNP{Hoflich95} and \citeNP{BaronHM96}) that the
Sobolev theory, developed for isolated lines, is in error when lines
populate wavelength space so densely that there is significant overlap of
the intrinsic line profiles in the co-moving frame. While there have been
no detailed studies of the magnitude of this error in the present context,
such an error undoubtedly occurs to some extent. Unfortunately, the very
great number of lines which the above discussion shows must be included in
a calculation makes a Sobolev treatment the only viable one with present
technology. An ``exact'' numerical treatment approaching the accuracy of
Sobolev theory would require several tens of frequency points per line.
Even allowing for significant overlap, this implies a number of
frequency grid points at least several times the number of lines---at the
very least, several million points. There are at present
no numerical techniques up to the task of representing the frequency
variation of the opacity with unambiguously sufficient
resolution. \citeN{EastmanP93} compared supernova spectra calculated with
and without the Sobolev approximation and found minimal differences.
Thus, while the objection is correct in principle, it is rendered moot in
current practice.

There is another indication that such a scattering-mediated
``thermalization'' may be taking place in \sneia. \citeN{HoflichKW95} and
\citeN{BaronHM96} note that the agreement of their calculated maximum-light
spectra with observations is significantly improved by the inclusion of
additional thermalization. They have attributed this to a lack of accurate
collision cross sections. Under this assumption, they empirically fit an
enhancement to the collision rates they employ, using the agreement of
their simulations with observed spectra as a criterion. This may well be an
appropriate exercise. On the other hand, the approximations they employ to
determine collision rates (the same as we have employed above) work well on
average in other astrophysical and terrestrial applications; in particular,
there is no reason to suspect that these approximations are {\em
systematically} incorrect other than the disagreement of computed supernova
spectra with observations.  The additional thermalization resulting from
splitting in a sufficiently large number of lines may be another, less
artificial, way to achieve this same effect.

\subsection{Frequency-Integrated Mean Opacities}\label{MeanOpac}

Simple analytic light curve models such as the one presented in
section~\ref{AnLCMod} and most, if not all, radiation hydrodynamic
calculations of Type Ia supernovae performed to date, rely upon the
existence of a well-defined, frequency-integrated mean opacity.  A natural
choice for this mean opacity is the Rosseland mean, which has a long
history of use in astrophysical contexts.  The formally correct choice is,
of course, the flux mean, but the whole point of using a mean opacity is to
avoid the multi-group calculation of the flux which is essential to the
computation of the flux mean in the first place!  In this section we use
the results of multi-group radiation transport calculations to compare the
Rosseland mean and flux mean opacities near maximum light in the
delayed-detonation model DD4 of \citeN{WoosleyW94}.

The second monochromatic radiation transport moment equation may be
written as
\begin{eqnarray}
&&{1\over{c^2}}{{DF_\nu}\over{Dt}} + {{\partial P_\nu}\over{\partial r}}
+ {{3P_\nu-E_\nu}\over r} +
{2 \over c^2}{v\over r} \left(F_\nu+4\pi N_\nu\right) + 
 \nonumber \\
&& {1\over c^2}{\partial v\over\partial r} \left(2 F_\nu-4\pi
N_\nu\right)
-{\partial\over\partial\ln\nu}
\left({v\over r}F_\nu + \left({\partial v\over\partial r} -
 {v\over r}\right) 4 \pi N_\nu\right)
 = -{1 \over c} \rho\kappa_\nu F_\nu
\label{monoradm}
\end{eqnarray}
where $N_\nu$ is the third moment of the radiation field specific intensity.
If this equation, as written, is integrated over frequency, one
obtains equation~\ref{radm}. The right hand side of equation~\ref{radm} can be
written as $-\rho\kappa_F F/c$, where the flux mean opacity,
$\kappa_F$, is defined as
\begin{equation}
\kappa_F \equiv F^{-1} \int_0^\infty \kappa_\nu F_\nu\, d\nu
\label{Fmeankappa}
\end{equation}
This is clearly the correct quantity to use for the mean opacity.  The problem,
as mentioned above, is that the calculation of the flux mean requires prior
knowledge of $F_\nu$ and thus requires solution of the frequency-dependent
problem. On the other hand, having calculated $F_\nu$ by some much more
complex calculation for a variety of models, we can try to discover
regularities in the behavior of $\kappa_F$ which may be useful in more
approximate calculations.

At large enough optical depth ($\tau_{thm}>>1$) the gas and radiation
field will be thermalized and $E_\nu\approx 4\pi B_\nu(T)$. The
radiation field will be isotropic, so that $P_\nu\approx E_\nu/3$.
The expansion is homologous so $v/r=\partial v/\partial r=
1/t$. Finally, since $\rho\kappa_\nu c t >> 1$, the time derivative
term in equation~\ref{monoradm}, $c^{-2}DF_\nu/Dt$, may be set to
zero. Substituting these relations into equation~\ref{monoradm} gives
\begin{equation}
{\partial F_\nu\over\partial \ln\nu} - (3 + \rho\kappa_\nu c t)F_\nu =
{4\pi c t \over 3}{\partial T\over \partial r}
{\partial B_\nu(T)\over\partial T}
\end{equation}
where we have written $\partial B_\nu(T)/\partial r=\partial
B_\nu(T)/\partial T\times \partial T/\partial r$. This equation is
straightforward to solve:
\begin{equation}
F_\nu = -{4\pi ct\over 3}{\partial T\over \partial r}
\int_\nu^\infty
{\partial B_\nup(T)\over\partial T}\left({\nu\over\nup}\right)^3
\exp\left(-ct\int_\nu^\nup\nupp^{-1}\rho\kappa_\nupp\,
d\nupp\right) 
\nup^{-1}\, d\nup
\label{FnuRos}
\end{equation}
The Rosseland mean opacity can now be determined from the definition
\begin{equation}
F={-4\pi\over 3 \rho\kappa_R}
{\partial T\over \partial r}
\int_0^\infty{\partial B_\nu(T)\over\partial T}\, d\nu.
\label{Fbol}
\end{equation}
Integrating equation~\ref{FnuRos} over $\nu$ and combining with 
equation~\ref{Fbol} gives
\begin{equation}
\kappa_R={\rho\pi t\over aT^3}
\int_0^\infty d\nu
\int_\nu^\infty
{\partial B_\nup(T)\over\partial T}\left({\nu\over\nup}\right)^3
\exp\left(-ct\int_\nu^\nup\nupp^{-1}\rho\kappa_\nupp\,
d\nupp\right) 
\nup^{-1}\, d\nup.
\label{kapparos}
\end{equation}
Note that we have used a large number of approximations to derive this
expression: a large optical depth, a thermalized radiation field, and a weak
time dependence of the radiation field. The most important of these is that
the radiation field be thermalized. We have shown above that we do not
expect this to be the case; we demonstrate in the following that the
radiation field is in fact not thermalized.

We have undertaken a series of light curve calculations with a more modern
version of the code EDDINGTON described in \cite{EastmanP93}.  Here we
present the result of just one such calculation---Paper II will contain
a more systematic examination of the light curves predicted
for a variety of SNe~Ia models.

Since our purpose was to make illustrative calculations
and not to model any individual supernova, the size of the calculations was
kept small enough to run on a workstation in reasonable time. Thus, we have
assumed LTE and employed a frequency grid with 3500 points from 33\AA\ to
$45\mu$, using the monochromatic opacity approximation given by
equation~\ref{kappa_exp} to represent all but the strongest 3000 lines in the
spectrum.

In EDDINGTON, the gas temperature is determined by solving the
time-dependent first law of thermodynamics. Heating is from radiative
absorption, fast particles from Compton scattering of gamma-rays, and decay
positrons. Losses are from expansion and radiative emission.  Near peak,
because the energy density is dominated by radiation the same gas
temperature would be obtained by balancing heating and cooling and ignoring
the gas pressure contribution to $PdV$ losses; clearly, however, the $PdV$
work done by the radiation must still be included. The local energy
deposition from radioactive decay is determined with reasonable accuracy,
especially before 50 days, by doing a separate deterministic transport
calculation for each $\gamma$-ray line emitted in the decay cascade from
\nifsx through \fefsx as described in \cite{WoosleyEWP94}. Energy from
\cofsx positrons, irrelevant to the light curve near peak, was deposited
{\it in situ}.  \citeN{BlinnikovBPEW96} obtain excellent agreement in
results on test problems run with EDDINGTON and with the multi-group
implicit radiation-hydrodynamics code STELLA (\citeNP{BlinnikovB93},
\citeNP{BartunovBPT94}) when the same approximations are used in both codes
for the line opacity.

Because the probability of photon destruction from ``line splitting'' is so
large, the best that an LTE calculation can do to represent this inherently
non-LTE effect is to redistribute the photon energies thermally. Thus the
line opacity of the 3000 strongest lines was taken to be purely absorptive.
The opacity from the remaining $3\times10^5$ lines treated by the opacity
approximation of \cite{EastmanP93} was taken to be purely scattering.
While this is a crude approximation to the photon destruction described
above, it is the best one can do without performing an extensive
time-dependent NLTE calculation, one that is beyond reach of present
computing resources.

\begin{figure}[tp]
\postfig{l}{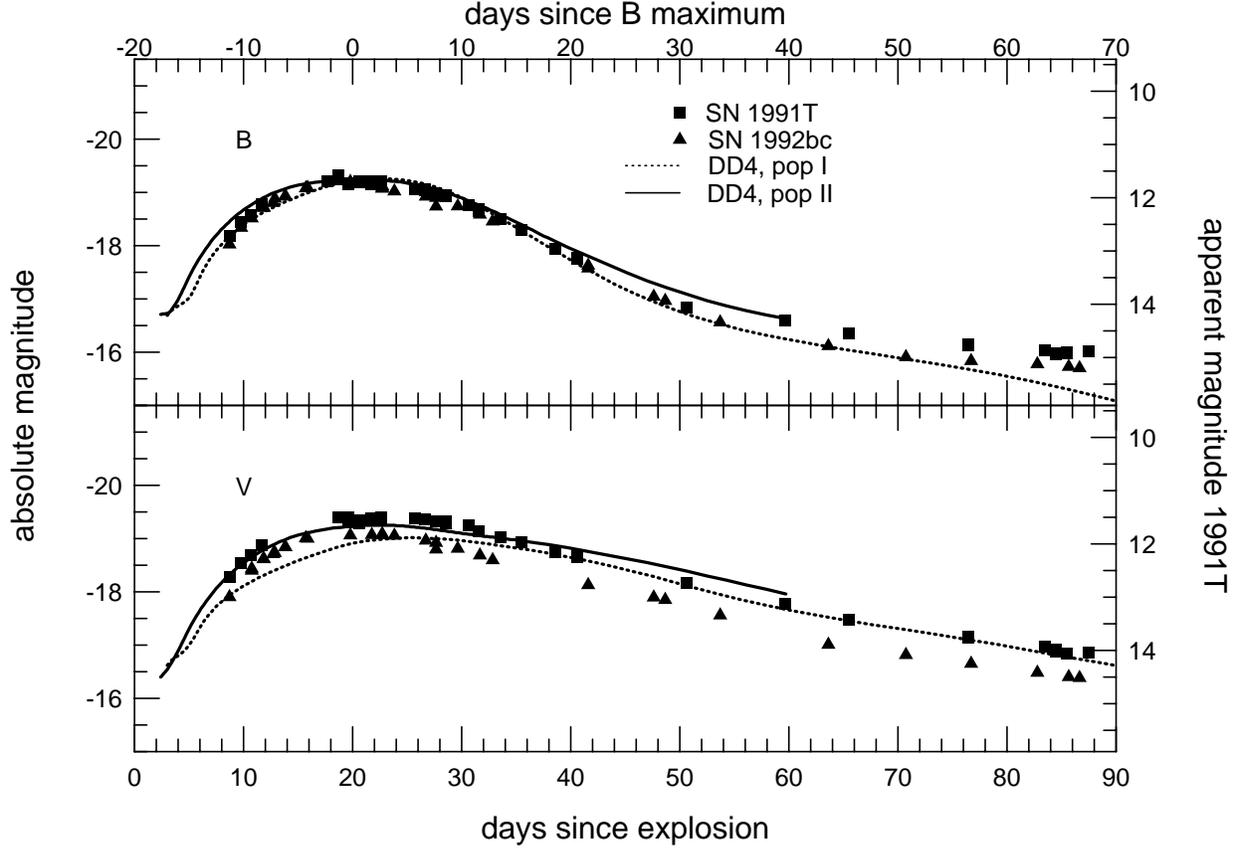}
\caption{B and V band light curves of delayed detonation Model DD4 of
\protect\citeN{WoosleyW91}, calculated as described in the text, compared
with data from SN 1991T and SN 1992bc (data from \protect\citeNP{Hamuyetal96}).
The solid line was calculated from the model with
population~II primordial abundances, the dotted line with population~I.  The
population I model, though too blue in $B-V$, gives the observed bolometric
risetime of $\sim19$ days, while the bolometric light curve of the
population II model rises too quickly, in 15 days.
}
\label{light curve}
\end{figure}

Figure~\ref{light curve} shows the B and V band light curves of model DD4, as
computed by this LTE prescription, compared with data from observations of
SNe 1991T and 1991bc. While not an exact reproduction of either object, the
fit is relatively good in B and V. In both
cases we have taken the basic explosion model, computed from a pure C/O
white dwarf, and added population II ($10^{-2}\times$ solar) and population
I (solar) abundances. For the population~II model,
the bolometric curve reaches maximum in only 15 days.  The
population~I model, however, reproduces the observed bolometric risetime of
18 days \cite{VaccaL96b}, though it gives a somewhat worse fit to the
observations in V.  These curves are remarkable as they are the first
models of \sneia light curves which peak as late as supernov\ae\ are observed
to do (c.f. \citeNP{HoflichKW95}).  For the first sixty days, the agreement
with observation is reasonably good. After this, the supernov\ae\ are
observed to make a transition to a quasi-nebular, emission-line spectrum,
one which bears little resemblance to a blackbody at {\em any} temperature.
An LTE calculation would not be expected to be a reasonable representation
of the dominant physics; we suspect that any agreement from this model
at such a late stage is therefore entirely fortuitous.

\begin{figure}[tp]
\postfig{p}{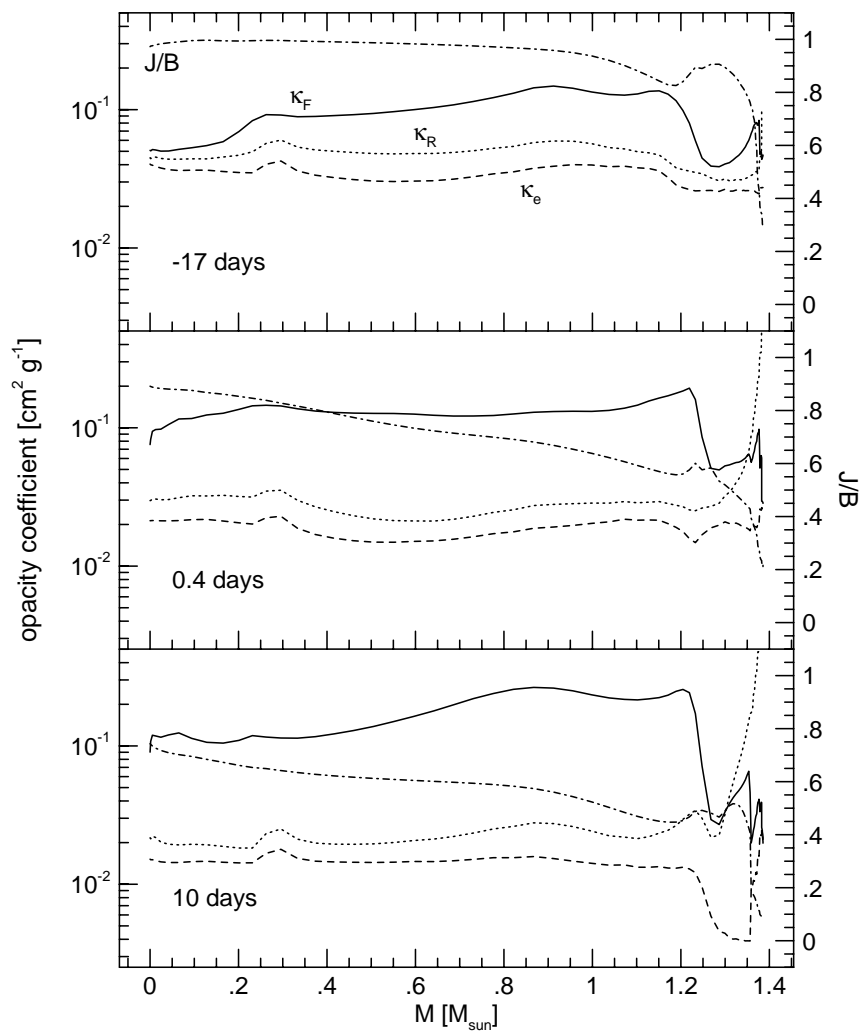}
\vspace{0.5cm}
\caption{Comparison of flux mean ($\kappa_F$), Rosseland mean
($\kappa_R$) and electron scattering ($\kappa_e$) mass opacity
coefficients in Model~DD4 versus Lagrangian mass coordinate at three
times before and after bolometric maximum. Also shown is the ratio of
the frequency integrated radiation field mean intensity to the Planck
energy density at the local gas temperature ($J/B$). Where $J/B<1$,
the gas and radiation field are not in equilibrium.}
\label{kappaFRE}
\end{figure}

Figure~\ref{kappaFRE} displays the run of $\kappa_F$, $\kappa_R$ and
$\kappa_e$ (the electron scattering mass opacity coefficient), at three
different times after explosion. At all times the flux mean opacity is
substantially greater than the Rosseland mean, by as much as a factor of 10
near maximum light. The effect this would have on the computed bolometric
light curve is comparable to a factor of three change in the total mass of
the star (when $E/M$ is held constant)! We have found similar results
comparing $\kappa_F$ and $\kappa_R$ from calculations of other explosion
models.

The reason that $\kappa_F$ is so different from $\kappa_R$ is
explained in part by the steep frequency dependence of the line
opacity at UV wavelengths ($d\ln\kappa_\nu/d\ln\nu\sim10$). Photon
transport takes place out on the Rayleigh-Jeans tail of the Planck
spectrum, and the steep frequency dependence of $\kappa_\nu$ means
that the mean opacity is very sensitive to that long wavelength slope.

In light of the discussion in section~\ref{PhotCollSec} we would expect
$F_\nu$ to be different than the diffusion flux given by
equation~\ref{FnuRos}. However, for the above calculation we assumed that the
strongest lines were purely absorptive. This should have the effect of
increasing the thermalization optical depth, driving the radiation field
and gas into thermalization, and therefore making the flux and Rosseland
mean opacities more equal in value.  With a more realistic NLTE treatment
of the line opacity, therefore, we would expect an even larger discrepancy
between $\kappa_F$ and $\kappa_R$.

Even at maximum light the departure from radiative equilibrium can be
significant.  Also shown in Figure~\ref{kappaFRE} $J/B=$ the ratio of
the frequency integrated mean intensity to $acT^4/4\pi$. At maximum
light the criterion that the radiation field be thermalized, required
in the derivation of the Rosseland mean opacity, is violated.  Where
$J/B < 1$, the color of the radiation field is hotter than a blackbody
at the same energy density. The Rossleand mean, which is weighted by
$\partial B_\nu/\partial T$ at the local gas temperature, samples the
opacity at longer wavelengths than does the flux mean. Because the
opacity falls off so strongly with increasing wavelength, the
resulting Rosseland mean is lower than the flux mean.

We can understand the variation of the mass opacity coefficients in
Figure~\ref{kappaFRE} as a combination of several effects.  From the
discussion above, we expect the extinction coefficient to be nearly
constant with radius, given roughly by the spectral density of lines
at some typical energy. As the density drops with radius, we expect
the mass opacity coefficient to rise proportionately. At ten days past
maximum, for example, the factor of three rise in $\kappa_F$ in going
from 0.5~\Msun\ to 1.0~\Msun\ is due almost entirely to the density
dropping by the same factor.  The constancy of $\kappa_F$ in the
central panel thus implies that the extinction coefficient has fallen
by about a factor of three in the same range. Because the flux mean
photospheric radius has by this time receded to $\sim0.68$~\Msun, this
change has little effect upon the luminosity.

The drop toward the center in the first panel is due to the higher
ionization of the iron group there. In the inner 0.3\Msun, the typical
ionization stages are Co~VII and VIII, while outside of this the
typical ions are Co~V and VI. The higher ionization stages possess fewer
strong transitions in the near UV, so that the spectral density of
lines, and hence the extinction coefficient, is smaller.

The smaller change in $\kappa_R$ with radius is due to the fact that as one
moves out, the temperature drops. This leads to a longer effective wavelength
in the weighting function $\partial B_\nu/\partial T$, where the monochromatic
opacity is lower, and a lower extinction coefficient. Since the density is
dropping as well, this leads to a smaller variation in $\kappa_R$.

It is interesting to note that if a constant opacity roughly equal to 
the Rosseland mean in the top panel is used in a grey solution of the
transport equations, the bolometric light curve peaks at $\sim12$ days,
not very different from the rise-times found by \citeN{HoflichKW95} for
similar explosion models, employing a Rosseland mean opacity.

\begin{figure}[tp]
\postfig{l}{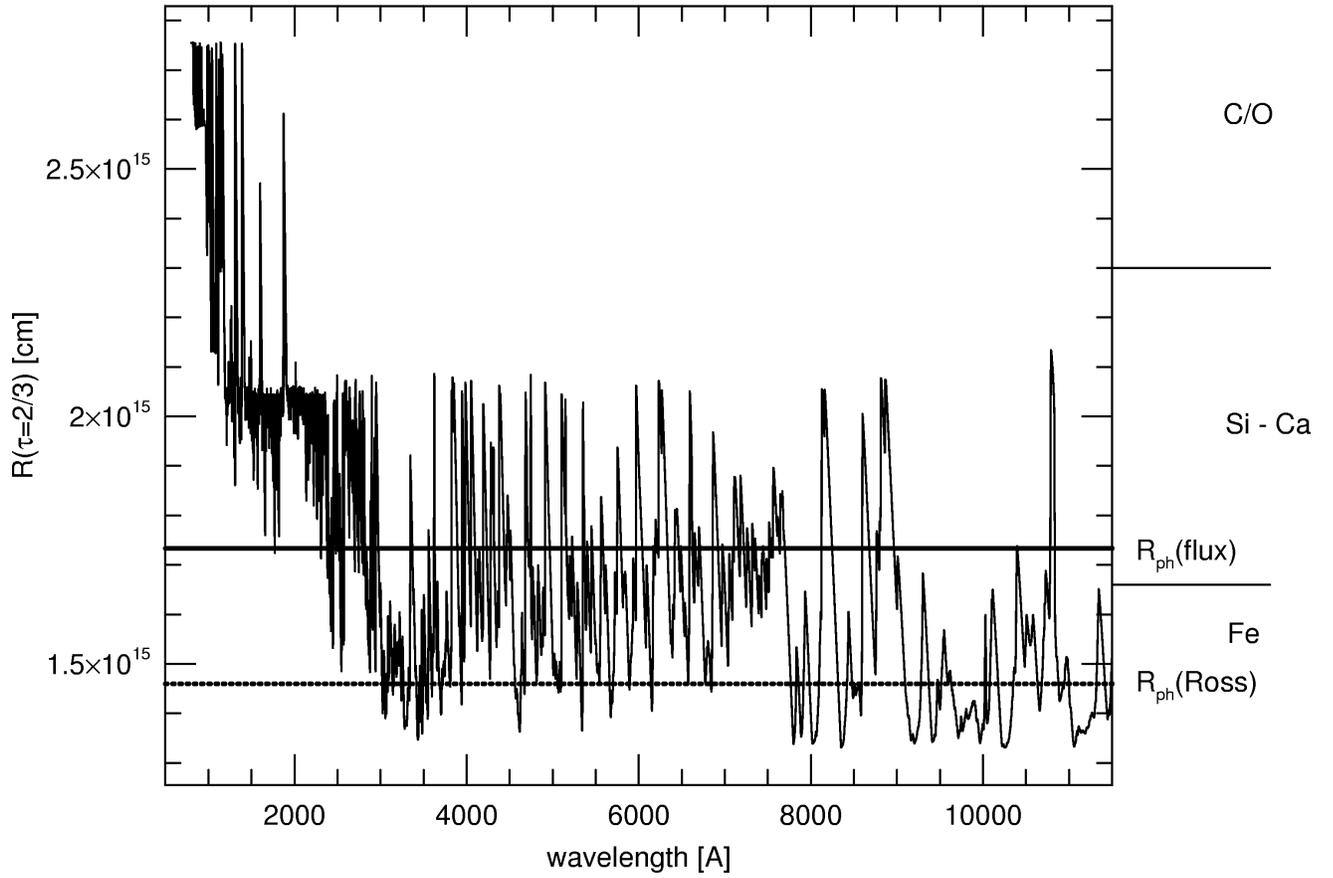}
\caption{This figure shows the photospheric radius as measured in the
observer frame at disk center for Model~DD4 at 14~days after explosion,
versus wavelength.
}
\label{rphotplot}
\end{figure}

Figure~\ref{rphotplot} displays the photospheric radius ($\tau=2/3$
surface) as a function of wavelength as measured in the observer frame
at the center of the disk, 14~days after explosion, with population II
primordial abundances. The photosphere is a very strong function of
wavelength whose behavior is dominated in the optical by a
considerable number of strong lines. Also shown are the Rosseland and
flux mean photospheric radii. Clearly, choosing either one as {\em
the} photosphere and assuming the emergent flux is a blackbody at some
temperature is a rather poor representation of the actual behavior. It
is interesting to note, however, that the flux mean photosphere at
this time lies at the outer extent of the iron-rich region as one
would expect from the lack of strong iron lines in the maximum-light
spectrum.

It is worth commenting that, whereas the assumption of constant
$\kappa$ made in Section~2 in the derivation of the analytic light
curve model may at first seem na\"ive, the results of the multi-group
calculations (Figure~\ref{kappaFRE}) show that it is, in fact, not
such a bad approximation, especially if a constant-density explosion
model is employed in the analytic treatment (it is actually
$\rho\kappa$ which is constant). This is fortunate, as it gives us
confidence that the simple light curve model described above gives a
fair representation of the response to changes in various parameters
of the explosion model.

\section{Time Dependence}

As a final, more technical issue, we can use the analytic solution we
have developed to examine the validity of various approximations to
the solution of the radiative transfer problem in supernov\ae.  While
it may seem obvious that \sneia are not steady state phenomena,
several papers have appeared in the literature in which the absolute
luminosity of some \sneia have been estimated ignoring the basic time
dependence of the transport physics.  This is a natural and indeed
necessary assumption for the calculation of NLTE maximum-light
spectra, as general, time dependent, NLTE calculations are beyond
current computational capabilities at present. It is, however,
important to understand the magnitude of the errors which may be
incurred by such approximations.

Among the most important and commonly-used approximations is that of
{\sl steady state}---that one or more of the time-dependent terms in
equations \ref{rade} and \ref{radm} can be ignored. These are
enormously attractive approximations as they greatly reduce the
computational complexity. A time independent problem is much easier to
solve than a time dependent one!

Steady-state amounts to the assertion that heating by energy deposition and
cooling by radiative processes (and perhaps by expansion as well) are
balanced {\em at all times}. There are two versions of this
approximation. The first asserts that the Lagrangian derivative (the first
term in equation \ref{rade}), is small compared with the flux
divergence. In this approximation, the supernova is no different than a
static stellar atmosphere. The second (employed, for example, by
\citeNP{NugentBHB95} and references therein), takes into account energy
loss from $PdV$ work but asserts that the Eulerian derivative $\partial
E/\partial t$ is negligible.

\begin{figure}[tp]
\postfig{l}{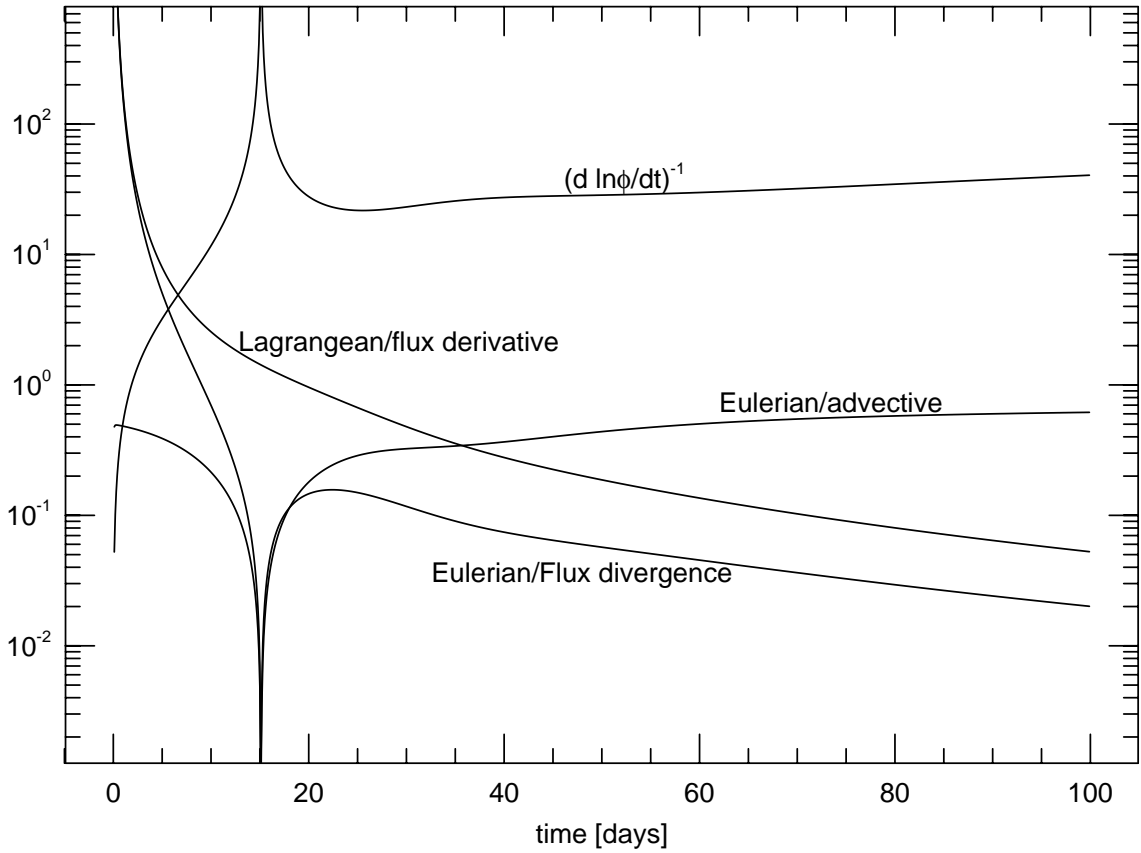}
\caption{The ratios of various terms in the transfer equation as functions
of time. Because a logarithmic scale is used, the absolute value of each
curve is plotted. Thus, at maximum (near 15 days) the Eulerian derivative
changes sign from positive to negative.}
\label{timeratios}
\end{figure}

In Figure \ref{timeratios} we present ratios of various terms in the
transfer equations as functions of time for our fully time dependent
solution. For clarity, only the fundamental mode is considered; the
inclusion of higher modes will make the time dependence different at
different depths in the ejecta but does not alter the character of the
solution nor the order of magnitude of the terms.

The ratio of the Lagrangian derivative to the flux divergence in
equation \ref{rade} is
\begin{equation}
{{D E}\over{Dt}}\left[
-{1\over{r^2}}{{\partial}\over{\partial r}} \left(r^2F\right)
\right]^{-1}
=
\psi
\left\{
 -1 
+ {{\tau_d}\over{\alpha}} {{\tilde{\epsilon}\theta\Lambda}\over{\phi}}
- {{4\tau_d}\over{t_{sc}\alpha}}
         \left( 1 + t/t_{sc} )\right)^{-2}
\right\}
\label{ratio1}
\end{equation}
Using the asymptotic value equation~\ref{asymphi},
equation~\ref{ratio1} becomes the ratio of the (current) diffusion
time to the expansion (elapsed) time and thus goes to zero in the
limit of large $t$. At late times, energy is deposited in the ejecta
and is immediately radiated away; thus the flux divergence must equal
the deposition, and it is appropriate to use a ``steady-state''
solution which balances instantaneous luminosity against the time
varying deposition rate, $\theta(t)\Lambda(t)$.

At peak light, $\dot{\phi}=0$ in equation \ref{inhomog} and in
equation~\ref{asymphi} as well, so equation~\ref{ratio1} becomes the
ratio of the present diffusion time to the elapsed time. Since peak
occurs when the diffusion time approximately equals the elapsed time,
the ratio in equation~\ref{ratio1} is $\sim 1$, implying that the time
dependent terms cannot be ignored. Thus while steady-state is a fine
approximation at late times it is quite a poor approximation at
earlier phases of the light curve. Indeed, it is only after 246 days
that the ratio in equation~\ref{ratio1} falls below 1\%.

In the second approximation, what one might call ``quasi-steady-state'',
only the Eulerian derivative is neglected. We can formulate the ratio of
the Eulerian to the advective derivative as
\begin{equation}
{{\partial E/\partial t}\over{(v\cdot\nabla)E}} = 
{{t_{sc}}\over{4}}\left[{{\tilde{\epsilon}\theta\Lambda}\over{\phi}}
- {{\alpha}\over{\zeta(t)/t_d}}\right](1+t/t_{sc})^2.
\label{ratio2}
\end{equation}
Once again, this ratio goes to zero in the limit of large time, but it also
goes to zero at peak, when $\dot{\phi}=0$. That this must be so is obvious
from examination of the light curves of the previous sections. Before peak,
the luminosity of the supernova is less than the rate at which energy is
being deposited in the ejecta, even accounting for losses due to
expansion. Shortly after peak, the luminosity is greater than the
deposition rate. This means that before peak, a store of energy is being
built up in the supernova and therefore that $\partial E/\partial t > 0$. After
peak, this ``excess'' energy is radiated away, and the energy loss is
greater than the loss due to expansion alone so that $\partial E/\partial t
< 0$. Since the sign of the Eulerian derivative changes when the supernova
traverses peak light, there must be a time near peak at which it is
zero. This does not mean that the term can be ignored, however. When the
effects of all modes are included, the time at which $\partial E/\partial t
= 0$ is different for each radius. As well, the derivative's value is
changing rapidly; one can see from the figure that only a few days before
and after maximum it is 30\% as great in magnitude as the advection term.

We wish particularly to draw attention to the erroneous conclusions drawn
in this regard by \citeN{BaronHM96}. In that work, its authors express
$\partial E/\partial t$ as a finite difference over a time interval $\delta
t$. They then go on to show that inclusion of this term has the effect of
an additional source or sink of radiation. While this is correct, when
comparing the magnitude of various terms they let $\delta t$, which would be
the ``time step'' in a finite difference treatment of the time dependent problem,
be the {\em elapsed} time. They then conclude that the term is small and
can be ignored. This is wrong. While it is obvious that in an implicit
difference scheme one approaches some sort of steady state if a sufficiently
long time step is employed, it is equally obvious that such a state need have
little semblance to a correct solution of the time-dependent
equations. This is especially so in that the solution at peak is not an
asymptotic limit. If they had chosen a time step small enough to preserve
accuracy in the finite difference, $\delta t$ would have been almost two
orders of magnitude smaller and their estimate of the relative size of the
Eulerian derivative would have been much the same as that determined here.

To see the error in the quasi-steady state approximation another way,
consider the limiting case of infinite opacity, radiative equilibrium, and
homologous expansion. Equations \ref{rade} and \ref{radm} then become
\begin{equation}
{{D E}\over{D t}} = -{{4\dot{R}}\over{R}}E.
\end{equation}
If we solve this equation directly, we find the expected result that
$E\propto t^{-4}$.  This clearly contradicts the notion that the intrinsic
time derivative of $E$ is zero!  In yet another demonstration, if we
consider the case in which we set $\dot{\phi}=0$, the light curve must
decline monotonically, following the energy deposition.  Thus the mere fact
that the light curve is observed to peak is testament to the error of the
quasi-steady-state.

\begin{figure}[tp]
\postfig{l}{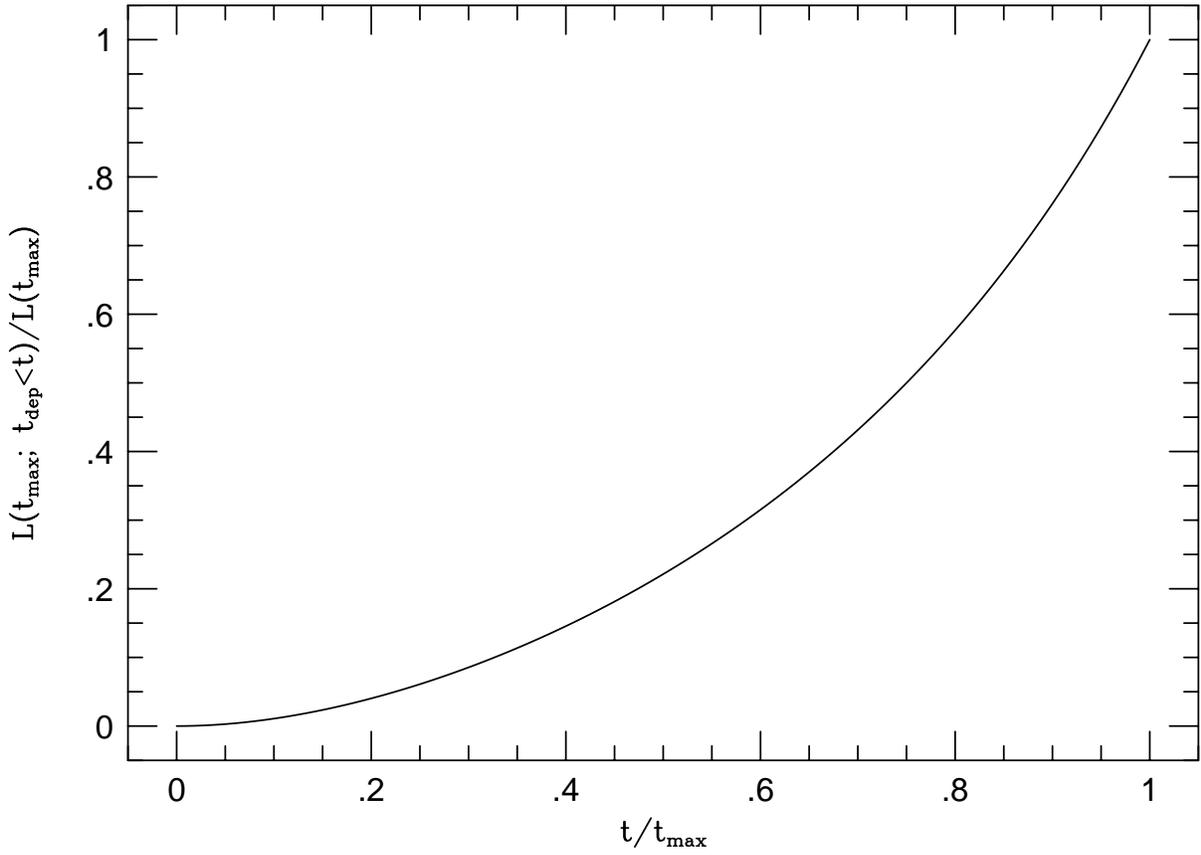}
\caption{The fraction of maximum light luminosity arising from energy
deposited before the time indicated on the abscissa.}
\label{oldphot}
\end{figure}

Perhaps the simplest way to gauge the effects of time dependence upon the
light curve is to examine when the luminosity emitted at a given time was
deposited in the ejecta. Figure \ref{oldphot} shows the cumulative fraction
of luminosity at maximum light as a function of deposition time for a
typical supernova model. It is clear that the ``residence time'' of the
energy which emerges from a supernova near maximum light is
significant. From the figure, for example, 50\% of the luminosity is energy
deposited at times earlier than 75\% the age of the supernova.

Energy is stored predominantly in the form of photons diffusing out through
the ejecta. Even the tactic of taking the thermal structure of the matter
from a light curve calculation, placing it in an atmosphere code, and
calculating the resulting spectrum does not do justice to the presence of
these ``old photons''. There is no reason to suspect that the radiation
temperature is the same as the matter temperature and thus without taking
some measure of the radiation field energy density from the light curve
calculation there is no way to characterize the spectrum or intensity of
these photons in the spectrum calculation. The only way to avoid a serious
omission in the physics of spectrum formation is to provide some measure of
the spectral shape and intensity of the ``old photons'' to the atmosphere
code. In practice, this means that the spectrum formation problem {\em is}
the light curve problem; one cannot avoid the inherent time-dependence in
calculating either light curves or spectra. Especially when attempting to
calibrate \sneia as cosmological distance indicators, it is necessary to
include all of these important physical effects.

It is true, however, that $\partial E/\partial t$ becomes small near
maximum light, and it may be true, despite the presence of ``old photons'',
that a ``quasi-steady-state'' treatment may not be too much in error at
this epoch.  Without a benchmark time-dependent NLTE calculation with which
to compare, however, it is difficult to assess the magnitude of any error
in the resulting luminosity. Since the spectrum-forming layers of the outer
atmosphere are primarily scattering, and since the shape of the
pseudo-continuum may have little to do with the gas temperature or the
luminosity as demonstrated above, a good agreement with the observed
spectrum {\em shape} may not in fact imply a reliable estimate of the
luminosity.

\section{Conclusions}

Radiation transport in SNe Ia has long been recognized to be fundamentally
different in character from that in other astrophysical objects, a
situation arising from the coincidence of large heavy element abundances
with a large velocity gradient and a diffusion time comparable with the
evolution time. The key to understanding this transport is the realization
that diffusion downward in frequency is as important to the escape of
energy as its diffusion outward in radius.  This coincidence arises due to
the rapid decline of the spectral density of optically thick lines with
increasing wavelength.  While the spectral shape of the trapped radiation
field is nearly Planckian at temperatures near $2\times10^4$~K, the flux
mediating this escape occurs predominantly at much lower energies.

One can ultimately do justice to such a complex process only by performing
detailed, time-dependent NLTE computations, those which are probably beyond
the present state of the art, or at least present computers. We have
demonstrated by more approximate means, however, some of the most important
effects.  For the present, the key to understanding lightcurve systematics
is to provide a simple model for the lightcurve, and the key
ingredient in this effort is to determine a mean opacity. The Rosseland
mean, so useful in static atmospheres, is significantly misleading in the
present context, leading to an incorrect picture of the escape of radiation
and serious errors in photospheric radii and temperatures.  We have shown
that the flux mean as determined by a multi-frequency computation is
remarkably constant in radius and with time during the peak phase of the
lightcurve. This allows one to make some progress by adopting a narrow
range of values for $\kappa_F$ in simple, constant-opacity models.

We have derived an analytic solution of the co-moving frame transport
equations which closely reproduces the results of more complex numerical
solutions. This allows one to examine a number of key features of the
lightcurve physics itself and of the observed systematics in \sneia data.
One result is the demonstration that the lightcurve and spectrum problem is
inherently time dependent and cannot be approximated by time-independent
calculations before at least 70 days past explosion.
Because the spectrum near maximum
light is formed by scattering lines atop a pseudo-continuum of line blends,
the shape of the spectrum is relatively insensitive to the luminosity. We
suspect that any correlation of line widths and velocities with bolometric
luminosity arises from differences in the underlying explosions; the
demonstration that a given model reproduces the {\em shape} of an observed
spectrum does not therefore imply that such a model reproduces the {\em
luminosity} of the observed explosion. Such a correlation thus requires,
for now, an independent, observational calibration.

Using the analytic model, we have explored the effect of changes in a
variety of parameters on the resulting lightcurve. These include the
opacity, explosion energy, \nifsx mass and distribution, and total mass. Of
these, there is only one parameter which by itself can explain the observed
correlation of peak width and luminosity: the total mass. All others, when
varied individually, lead to {\em anti-correlations}. This does not
necessarily imply that the mass of the explosion is {\em the} controlling
parameter; there may be various combinations of parameters which, when
altered in concert, lead to the same behavior. For example, if the opacity
can be shown to be a strong function of the \nifsx mass then the behavior
of models at a single mass may be able to reproduce the PR.  The fact that
variations in so fundamental a property of the explosion as the total mass
{\em can} explain the observed behavior is, to say the least, suggestive.

\acknowledgments

This work was supported (PAP) by the National Science Foundation (CAREER
grant AST9501634), by the National Aeronautics and Space Administration
(grant NAG~5-2798) and by the US Department of Energy
(W-7405-ENG-48). Philip Pinto gratefully acknowledges support from the
Research Corporation though a Cottrell Scholarship and from the General
Studies Group at Lawrence Livermore National Laboratory.  This work was
begun at the European Southern Observatory, Garching bei M\"unchen, where
both PAP and RGE gratefully acknowledge support from the visitors
program. Finally, we wish to thank W.D. Arnett, E. Baron, W. Benz,
A. Burrows, P. H\"oflich, R.P. Kirshner, B. Leibundgut, M. Mamuy,
D. McCray, J. Spyromilio, T.A. Weaver, and S.E. Woosley for many helpful
discussions and comments throughout the development of this work.

\bibliographystyle{apj}
\bibliography{apjmnemonic,main}   

\end{document}